\documentclass[entropy,article,submit,pdftex,moreauthors,10pt,a4paper]{mdpi}

\firstpage{1} 
\articlenumber{x}
\doinum{10.3390/------}
\pubvolume{xx}
\pubyear{2017}
\copyrightyear{2017}
\externaleditor{Academic Editor: name}
\history{Received: date; Accepted: date; Published: date}
\graphicspath{{./figures/}}

\usepackage{amssymb}
\usepackage{comment}
\usepackage{subcaption}

\usepackage{soul,color}
\soulregister\cite7
\soulregister\ref7
\soulregister\pageref7

\newcommand{\Vfire}{V_T} 

\newcommand{\Wcrit}{W_C} 

\newcommand{\gcrit}{\Gamma_C} 

\newcommand{\rhocrit}{\rho_C} 

\newcommand{\set}[1]{\left\{#1\right\}} 
\newcommand{\avg}[1]{\left\langle#1\right\rangle} 
\newcommand{\labelfig}[2]{\vbox{\vbox to 0pt{\hbox to 0pt{\textbf{\textsf{#1}}}\vss}\hbox{#2}}}

\Title{Self-Organized Supercriticality and Oscillations in Networks 
of Stochastic Spiking Neurons}

\Author{Ariadne A. Costa $^{1,2}$, Ludmila Brochini $^{3}$ and Osame Kinouchi $^{4,}$*}
\AuthorNames{Ariadne de Andrade Costa, Ludmila Brochini and Osame Kinouchi}

\address{%
$^{1}$ \quad Department of Psychological and Brain Sciences, Indiana University, Bloomington-IN, 47405, USA\\
$^{2}$ \quad  Instituto de Computa\c{c}\~ao, Universidade de Campinas, 
Campinas-SP, 13083-852, Brazil\\
$^{3}$ \quad Departamento de Estat\'\i stica, IME, Universidade de S\~ao Paulo, S\~ao Paulo-SP, 05508-090, Brazil\\
$^{4}$ \quad Departamento de F\'\i sica, FFCLRP, Universidade de S\~ao Paulo, 
Ribeir\~ao Preto-SP, 14040-901, Brazil }

\corres{Correspondence: osame@ffclrp.usp.br}

\abstract{Networks of stochastic spiking neurons are interesting models in the area of Theoretical Neuroscience, presenting both continuous and discontinuous phase transitions. Here we study fully connected networks analytically, numerically and by computational simulations. The neurons have dynamic gains that enable the network to converge to a stationary slightly supercritical state (self-organized supercriticality or SOSC) in the presence of the continuous transition. We show that SOSC, which presents power laws for neuronal avalanches plus some large events, is robust as a function of the main parameter of the neuronal gain dynamics. We discuss the possible applications of the idea of SOSC to biological phenomena like epilepsy and Dragon-king avalanches. We also find that neuronal gains can produce collective oscillations that coexists with neuronal avalanches, with frequencies compatible with characteristic brain rhythms.}


\begin{document}


\section{Introduction}

Neuronal network models are extended dynamical systems that may 
present different collective behaviors or phases characterized
by order parameters. The separation regions between phases can be
described as bifurcations in the order parameters, or 
phase transitions. In several 
models of neuronal activity, the relevant phase change 
is a continuous transition from an absorbing silent state  
to an active state~\cite{Herz1995,Beggs2003,Kinouchi2006}. 
In such continuous transition we have 
a critical point (in general, a critical surface) where 
concepts of  universality classes and critical exponents (among others) are valid. At criticality, we observe avalanches of activity described
by power laws for their size and duration. Also, the
avalanche profile shows fractal scaling. Since the 
landmark  findings of Beggs and Plenz in 2003~\cite{Beggs2003},
these behaviors have been reported also in biological  
networks, see reviews ~\cite{Chialvo2010,Markovic2014,Hesse2015}.

The motivation for the idea that criticality is important to
understand  neuronal activity is not only empirical. Several works
have shown that there are advantages for a network to operate at
the critical state~\cite{Kinouchi2006,Beggs2008,Shew2009,
Massobrio2015}. However, it is not clear how  biological
networks tune themselves to the critical region that, in
a parametric space with $P$ free parameters, has a maximum of $P-1$ dimensions. 

 An important idea discussed in several papers is that, since criticality depends on the strength of the synapses (links) between the neurons, a homeostatic mechanism for dynamic synapses tunes the network toward the critical 
region. There are two main paradigms: self-organization of Hebbian synapses~\cite{Arcangelis2006,Pellegrini2007,Arcangelis2010,Arcangelis2012,Arcangelis2012b,Kessenich2016}, and self-organization of dynamic synapses~\cite{Levina2007,Levina2009,
Bonachela2010,Costa2015,Campos2017} following Tsodyks and 
Markran~\cite{Tsodyks1997,Tsodyks1998}.
With these synaptic mechanisms it is possible to achieve, or at least to approximate, a self-organized critical (SOC) state.

With a different approach, we have shown recently that dynamic
neuronal gains, biophysically linked to firing dependent
excitability of the Axonal Initial Segment
(AIS)~\cite{Kole2012,Ermentrout2001,Benda2003,
Buonocore2016}, can also lead to
self-organized criticality~\cite{Brochini2016}. This new 
mechanism is simpler than dynamic synapses because, for biological
networks with $N$ neurons, we have of the order of $10^4 N$
synapses \cite{Tang2001} -- and thus $10^4 N$ dynamic equations -- 	
\textbf{}but only $N$ equations for the neuronal gains. 

It has also been observed in Brochini \emph{et al.}~\cite{Brochini2016}
that, for achieving exact SOC, all papers in the literature
used a time scale $\tau$ for synaptic recovery proportional to $N$,
with $N \rightarrow \infty$.
The use of this non-local information ($N$), and a diverging
recovery time $\tau$, is not plausible 
biologically. A similar time scale $\tau$ is present in the 
neuronal gain recovery. When we use a biological 
range for $\tau$ that does not scale with $N$, we observe that the network turns out (slightly) supercritical, a phenomenon
which we called self-organized supercriticality (SOSC).
That is, both dynamic synapses and dynamic gains with fixed
$\tau$, which seems to be the reasonable biological assumption, 
presents SOSC instead of SOC.

Here we report for the first time an extensive study of 
the neuronal gain mechanism and SOSC. First, we
present new mean-field results for phase transitions
in a fully connected model of integrate-and-fire 
stochastic neurons with fixed gains. We find both
continuous and discontinuous phase transitions. Then, we introduce
a simplified gain dynamics depending only on the $\tau$
parameter, which also has a simple mean-field solution in the
case of the continuous transition and presents SOSC.
We compare this solution with extensive simulations
for different system sizes $N$ and values for $\tau$. 
Surprisingly, we found collective oscillations produced by the gain dynamics
that coexist with neuronal avalanches.

\section{ {The model}}

We consider a fully connected network composed of $i=1,\ldots,N$ 
discrete-time stochastic neurons~\cite{Brochini2016,Gerstner1992,Gerstner2002,Galves2013,Larremore2014}.
The synapses transmit signals from some 
\emph{presynaptic} neuron $j$ to 
a \emph{postsynaptic} neuron $i$ with  {synaptic strength $W_{ij}$.
The Boolean variable $X_i[t]\in \set{0,1}$  denotes whether neuron $i$ fired between $t$ and $t+1$ and $V_i[t]$ corresponds to its membrane potential at time $t$.  Firing $X_i[t+1]=1$ occurs with probability 
$\Phi(V_i[t])$, which is called the \emph{firing function}~\cite{Galves2013,Larremore2014,DeMasi2015,Duarte2014,Duarte2015,Galves2016}.}

 {If a presynaptic neuron $j$ fires at discrete time $t$, then $X_j[t]=1$.
This event increments by $W_{ij}$ the potential of every postsynaptic 
neuron $i$ that  have not fired at time $t$.}
The potential of a non-firing neuron
may also integrate an \emph{external stimulus} $I_i[t]$. 
Apart from these increments, 
the potential of a non-firing neuron decays at each 
time step towards zero by a 
factor $\mu \in [0,1]$, which models 
the effect of a current  {leakage.} 

 {The} neuron membrane potentials evolve as:
\begin{equation}
V_i[t+1] = 
  \left\{
     \begin{array}{lcl}
        \displaystyle
        0  &\quad& \hbox{if $X_i[t]= 1$,} \\
        \displaystyle
         \mu V_i[t] + I_i[t] + \frac{1}{N}
         \sum_{j=1}^{N} W_{ij} X_j[t] &\quad&
            \hbox{if $X_i[t] = 0$.}
     \end{array}
   \right.
\label{modeldiscrete}
\end{equation}

This is a special case of the general model from~\cite{Galves2013} 
with the filter function $g(t-t_s) = \mu^{t - t_s}$, 
where $t_s$ is the time of the last firing of neuron 
$i$~\cite{Brochini2016}. In contrast to standard IF neurons,
the firing is not deterministic above a threshold,  { but
stochastic. We also} have $X_i[t+1]=0$ if $X_i[t]=1$ (refractory period  {of one time step}).

The firing function $0 \leq \Phi(V) \leq 1$ is sigmoidal, 
that is, monotonically 
increasing. We also assume that $\Phi(V)$ is 
zero up to some \emph{threshold potential} $\Vfire$.
If $\Phi$ is the shifted Heaviside step function 
$\Phi(V) = \Theta(V - \Vfire)$, 
we have a deterministic discrete-time leaky integrate-and-fire
(LIF) neuron. Any other choice for $\Phi(V)$ gives a stochastic neuron.
 
In Brochini \emph{et al.}~\cite{Brochini2016} we have studied
a linear saturating function
with neuronal gain $\Gamma$ similar to that used 
in~\cite{Larremore2014}. Here we study the so called
rational function that does not have a saturating potential,
see~Fig.~\ref{Fig1a}:
\begin{equation}
  \Phi(V) = \frac{\Gamma (V-\Vfire)}{1+\Gamma(V-\Vfire)} \:
  \Theta(V-\Vfire)\:. 
  \label{e.ratio}
\end{equation}

Notice that we recover the deterministic LIF model 
$\Phi(V) = \Theta(V-\Vfire)$ when $\Gamma \rightarrow \infty$.
 {The use of the rational instead of the linear
saturating function is convenient and gives
some theoretical advantages, 
for example to avoid the
anomalous cycles-$2$ observed in~\cite{Brochini2016}. }

\section{ {Mean-field calculations}}

The network's activity is measured by the fraction 
$\rho[t]$ of firing neurons (or density
of active sites):
\begin{equation}
  \rho[t] = \frac{1}{N} \sum_{j=1}^N X_j[t]\:.
  \label{e.rhot}
\end{equation}

The density of active neurons $\rho[t]$ can be computed from the 
probability density $p[t](V)$ of potentials at time $t$:
\begin{eqnarray}
  \rho[t] = \int_{-\infty}^\infty \Phi(V) p[t](V)\, dV \:,
  \label{e.rho0}
\end{eqnarray}
where $p[t](V)\,dV$ is the fraction of neurons with potential 
in the range $[V,V+dV]$ at time $t$.

Neurons that fire between $t$ and $t+1$ have their 
potential reset to zero.  
They contribute to $p[t+1](V)$ a Dirac impulse at 
potential $V = 0$, with amplitude 
$\rho[t]$ given by equation~(\ref{e.rho0}).
The potentials of all neurons
also evolve according to equation~(\ref{modeldiscrete}). 
This process modifies $p[t](V)$ also for $V \neq 0$.

 In the mean-field limit, we assume that the synaptic weights $W_{ij}$ follow a distribution with average $W= \avg{W_{ij}}$ and finite variance. By disregarding correlations,  the term in 
 Eq.~(\ref{modeldiscrete}) corresponding to the sum of all presynaptic inputs simplifies to $W \rho[t]$.

If the external input is constant, $I_i[t] = I$, 
a stationary state is achieved, which depends only on
the average synaptic weight  {$W$}, 
the leakage parameter $\mu$ and
the parameters that define the function $\Phi(V)$, 
that is, $\Gamma$ and $\Vfire$. 
In Brochini \emph{et al.}~\cite{Brochini2016} it is shown
that the stationary $p(V)$ is composed of delta peaks
with height $\eta_k$ situated at voltages $U_k$ given by:
\begin{eqnarray}
U_0 &=& 0 \:, \label{e.u0}\\
U_k &=& \mu U_{k-1} + I +  W \rho\:, \label{e.uk}\\
\eta_k &=& \left( 1 - \Phi(U_{k-1})\right)\eta_{k-1} \:,\label{e.srhok}\\
\rho &= &\eta_0 = \sum_{k=0}^{\infty} \Phi(U_k) \eta_k\:,\label{e.seta0} 
\end{eqnarray}
for all $k\geq 1$. Here $U_k$ corresponds to the potential value of the population of neurons that have firing age $k$.  { {The firing age is the amount of time steps since the neuron fired for the last time.}}
The normalization condition $\sum_{k=0}^\infty \eta_k\;=\;1$ 
must be included explicitly. 
Equations~(\ref{e.uk}--\ref{e.seta0})
can be solved numerically for any firing function $\Phi$,
so this result is very general.

\section{Results}
\subsection{Phase transitions for the rational $\Phi(V)$}

In terms of non-equilibrium
statistical physics, $\rho$ is the order parameter,
$I$ is an uniform external field and $\Gamma$ and $W$
are the main control parameters. The activity $\rho$ also
depends on $\Vfire$ and $\mu$.

\subsubsection{The case with $\mu > 0, I=0, \Vfire = 0$}

\begin{figure}[!ht]
 \centering
   \begin{subfigure}{.5\textwidth}
  \centering
  \includegraphics[width=1.0\linewidth]{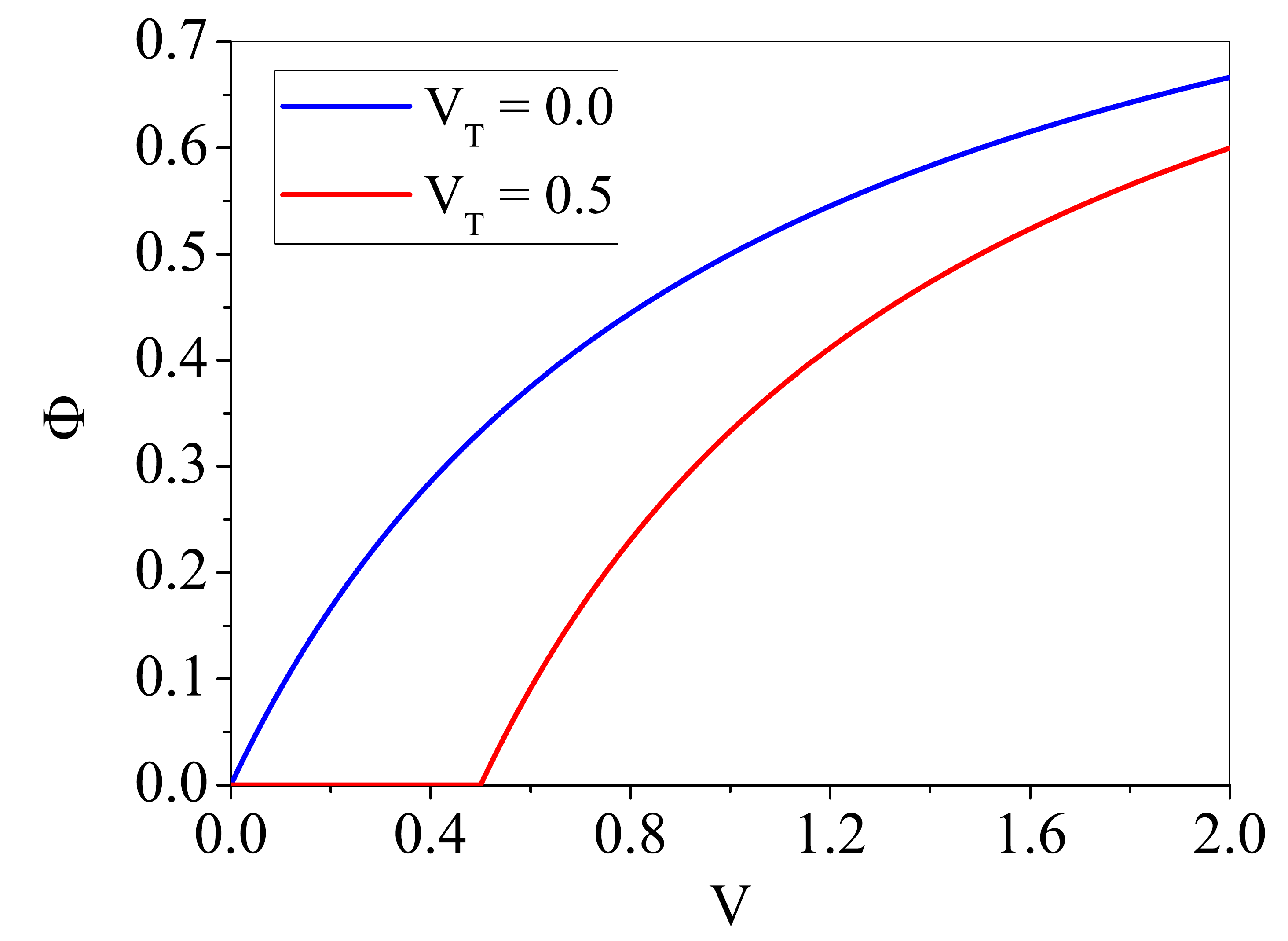}
  \caption{}
  \label{Fig1a}
\end{subfigure}
\begin{subfigure}{.5\textwidth}
  \centering
  \includegraphics[width=1.0\linewidth]{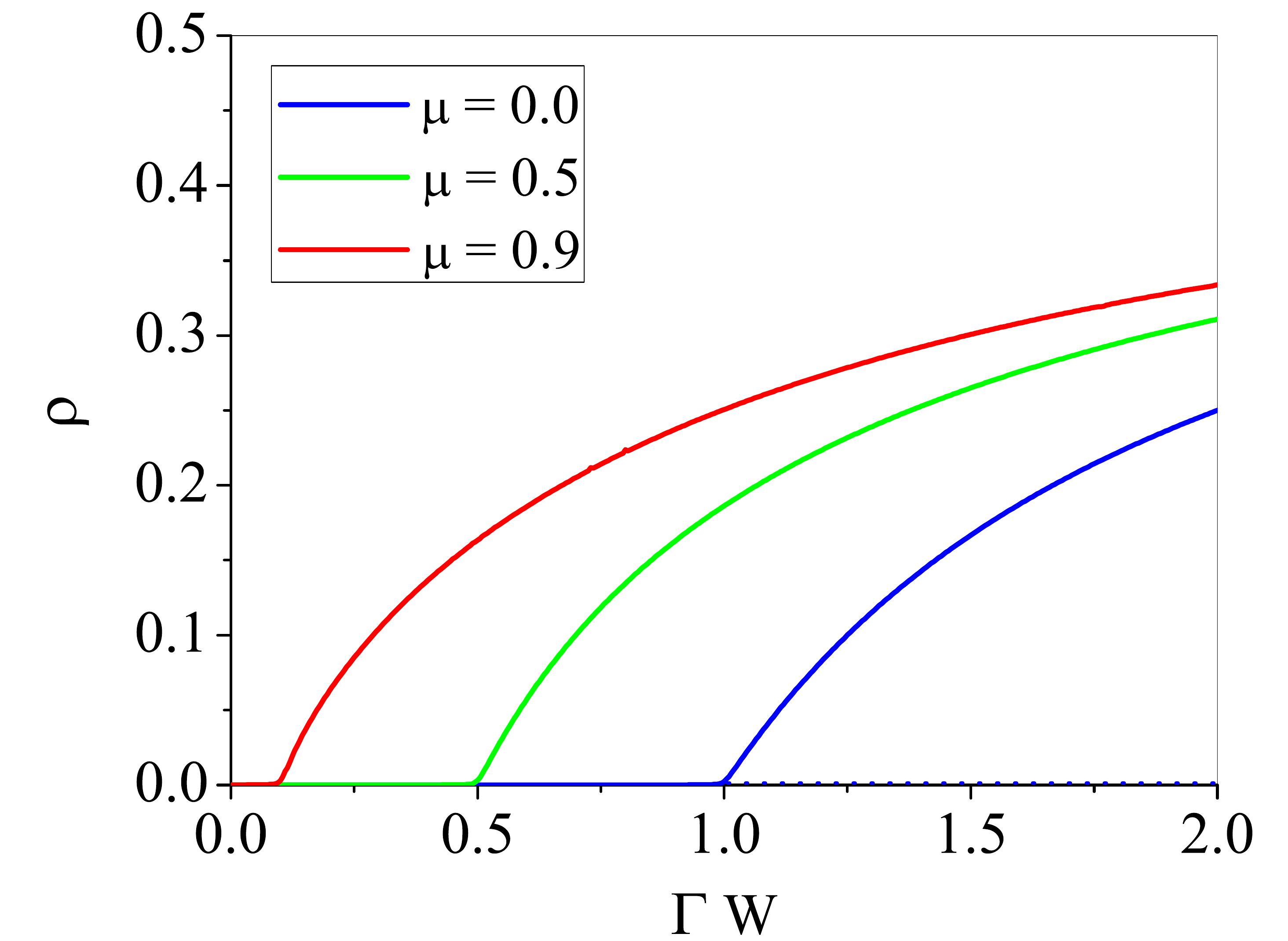}
  \caption{}
  \label{Fig1b}
\end{subfigure}\\
    \begin{subfigure}{.5\textwidth}
  \centering
  \includegraphics[width=1.0\linewidth]{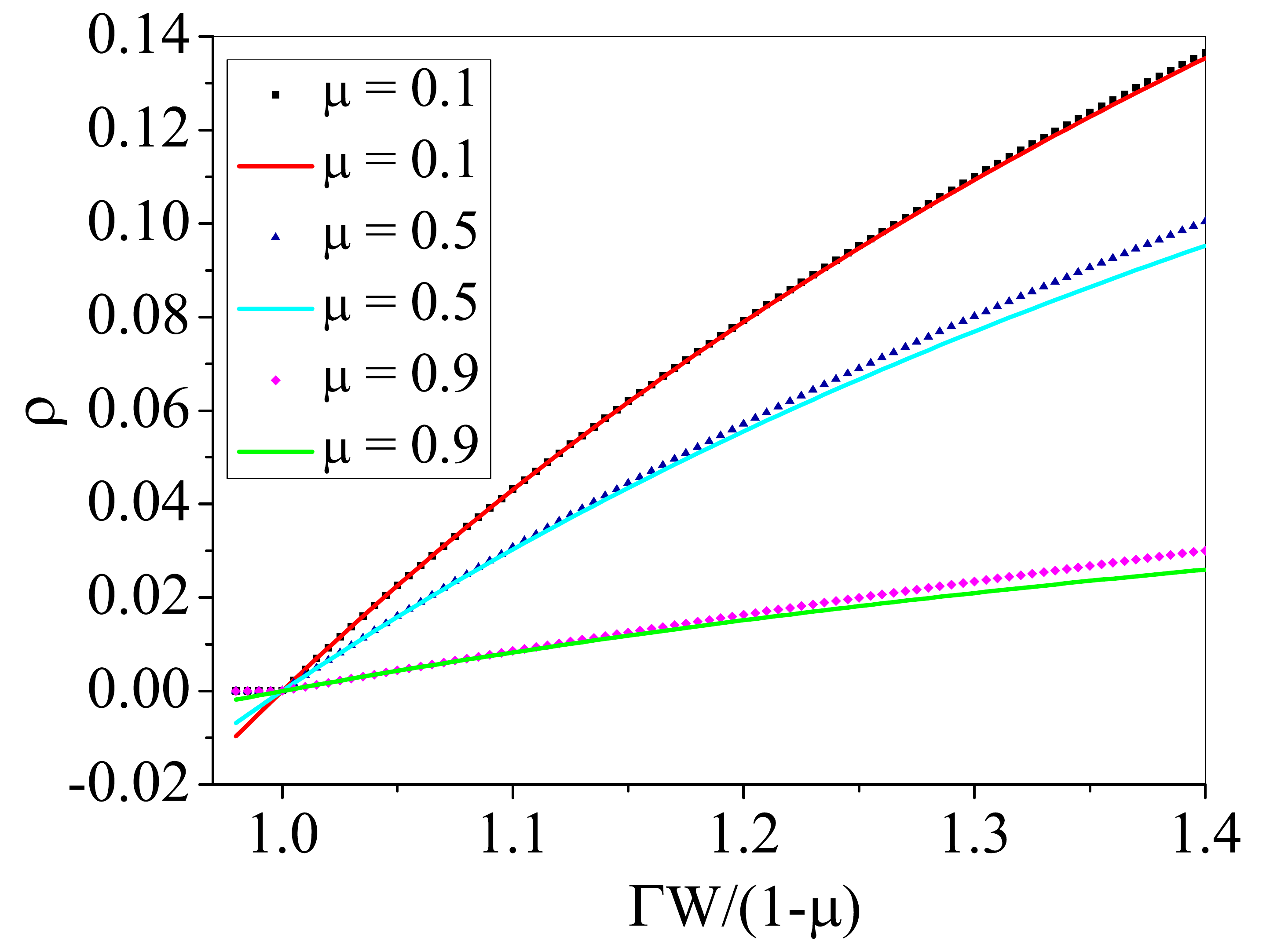}
  \caption{}
  \label{Fig1c}
\end{subfigure}

    \caption{{
    \bf Firing densities and phase diagram for $\bf \Vfire = 0,  {I=0}.$} 
      (a) Examples of the rational firing function
      $\Phi(V)$ for $\Gamma = 1,
    \Vfire = 0.0$ and $\Gamma =1, \Vfire = 0.5$;
  (b) Firing density $\rho(\Gamma W)$  for $\mu=0.0, 0.5, 0.9$. 
  The absorbing state $\rho_0=0$ looses stability after 
  $\Gamma W >\gcrit \Wcrit=1-\mu$; 
  (c) Comparison, near the critical region,
  between order parameter $\rho$ obtained numerically
  from Eq.~(\ref{e.seta0}) (points) and from the 
  analytic approximation Eq.~(\ref{rhoapprox}) (lines).  
  }
\label{GxW}
\end{figure}

 {By using Eqs.~(\ref{e.u0})-(\ref{e.seta0}),} 
we obtain numerically $\rho(W,\Gamma)$
for several values of $\mu > 0$, for the case
with $I =0, \Vfire = 0$ (Fig.~\ref{Fig1b}).  
Only the first $100$ peaks $(U_k,\eta_k)$ were considered,
since, for the given $\mu$ and $\Phi$, there was no significant 
probability density beyond that point. The same numerical method
can be used for studying the cases $I \neq 0, \Vfire \neq 0$.

We also obtained an analytic approximation (see Appendix)
for small $\rho$:
\begin{equation}
\rho \approx  \left(\frac{1}{ 2 +  \mu + \mu^2/(1-\mu)} \right)
\frac{\Gamma-\gcrit}{\Gamma}  
\: \propto \:  \Delta_\Gamma^\beta \:,\label{rhoapprox} 
\end{equation}
where $\gcrit = (1-\mu)/W$ defines the critical line
and $\Delta_\Gamma = (\Gamma - \gcrit)/\Gamma$
is the reduced control parameter.
So, the critical exponent  {for the order parameter near criticality} is $\beta = 1$,
characteristic of the mean-field directed percolation 
(DP) universality class  \cite{Hinrichsen2000}.
We also compare Eq.~(\ref{rhoapprox}) with the numerical results for
$\rho(\Gamma,\mu)$ in Fig.~\ref{Fig1c}.

\subsubsection{Analytic results for $\mu = 0$}

In the case $\mu =0$ it is possible to do a simple mean-field analysis valid for $N \rightarrow \infty$. This case is illustrative because it presents all phase transitions that occur with
$\mu > 0$.

When $\mu = 0$ and $I_i[t]=I$ (uniform constant input), 
the stationary density $p(V)$ consists of only
two Dirac peaks at potentials  { {$U_0 = 0$ and
$U_1 =  I + W \rho$}}. Eq.~(\ref{e.seta0}) simplifies to:
\begin{equation}
  \rho = \rho \Phi(0) + (1-\rho) \Phi(I + W \rho) \:, \label{e.rho20}
\end{equation}
since $\eta_0 = \rho$ and $\eta_1 = 1 - \rho$. 

By inserting  the function Eq.~(\ref{e.ratio}) in 
Equation~(\ref{e.rho20}), and remembering that $\Phi(0) = 0$, we get:
\begin{equation}
2 \Gamma  W \rho^2 - (\Gamma W + 2 \Gamma (\Vfire - I) - 1)\rho +
\Gamma (\Vfire - I) = 0\:,
\end{equation}
with solutions:
\begin{eqnarray}
\rho^\pm &=& \frac{\Gamma (W + 2 \Vfire - 2I) - 1 \pm
\sqrt{\Delta}}{4 \Gamma W}\:,\\
\Delta &=& (\Gamma (W + 2 \Vfire - 2I) - 1)^2
- 8 \Gamma^2 W (\Vfire - I)\:, \label{rhoWI}\\
\end{eqnarray}

\subsubsection{The case with $I =0, \Vfire = 0$: continuous transition}

For $V_T = I = 0$, we have:
\begin{eqnarray}
\rho(W) &= &\frac{1}{2} \left(\frac{W-\Wcrit}{W}\right)^\beta 
 = \frac{1}{2} \left(\frac{\Gamma-\gcrit}
{\Gamma}\right)^\beta \:, \label{rhocrit} 
\end{eqnarray}
where the phase transition line is
\begin{equation}
\gcrit = 1/\Wcrit \:, \label{PTL}
\end{equation}
and the critical 
exponent is $\beta = 1$. This corresponds to a standard mean-field
continuous (second order) absorbing state phase transition
(Fig.~\ref{Fig2a} and~\ref{Fig2b} with $\Vfire =I = 0$). 

\subsubsection{The case with $I < \Vfire$: discontinuous transition}
For $I < \Vfire  $, we have discontinuous (first order)
phase transitions when $\Delta = 0$, see Eq.~(\ref{rhoWI}):
\begin{equation}
\left(\gcrit \Wcrit + 2\gcrit (\Vfire - I) - 1 \right)^2 
= 8 \gcrit^2 \Wcrit (\Vfire - I)  \:,
\end{equation}
which, after some algebra, leads to the phase transition lines:
\begin{eqnarray}
\gcrit \Wcrit &= & \left(1+ \sqrt{2 \gcrit (\Vfire - I)} \right)^2\:,
\label{GW}  \\
\gcrit &=& \frac{1}{\left(\sqrt{\Wcrit} 
- \sqrt{2(\Vfire - I)}\right)^2}
\label{GW2}
\end{eqnarray}
which have the correct limit, Eq.~(\ref{PTL}), 
when $\Vfire \rightarrow 0, I \rightarrow 0$.
The transition discontinuity is:
\begin{equation}
\rhocrit = \frac{\sqrt{\Vfire-I}}{\sqrt{2\Wcrit}} 
=\frac{1}{2}\:\frac{\sqrt{2\gcrit 
(\Vfire-I)}}{1+\sqrt{2\gcrit(\Vfire-I)}}\:. 
\label{rhocritD}
\end{equation}

\begin{figure}[!ht]
 \centering
  \begin{subfigure}{.45\textwidth}
  \centering
  \includegraphics[width=1.0\linewidth]{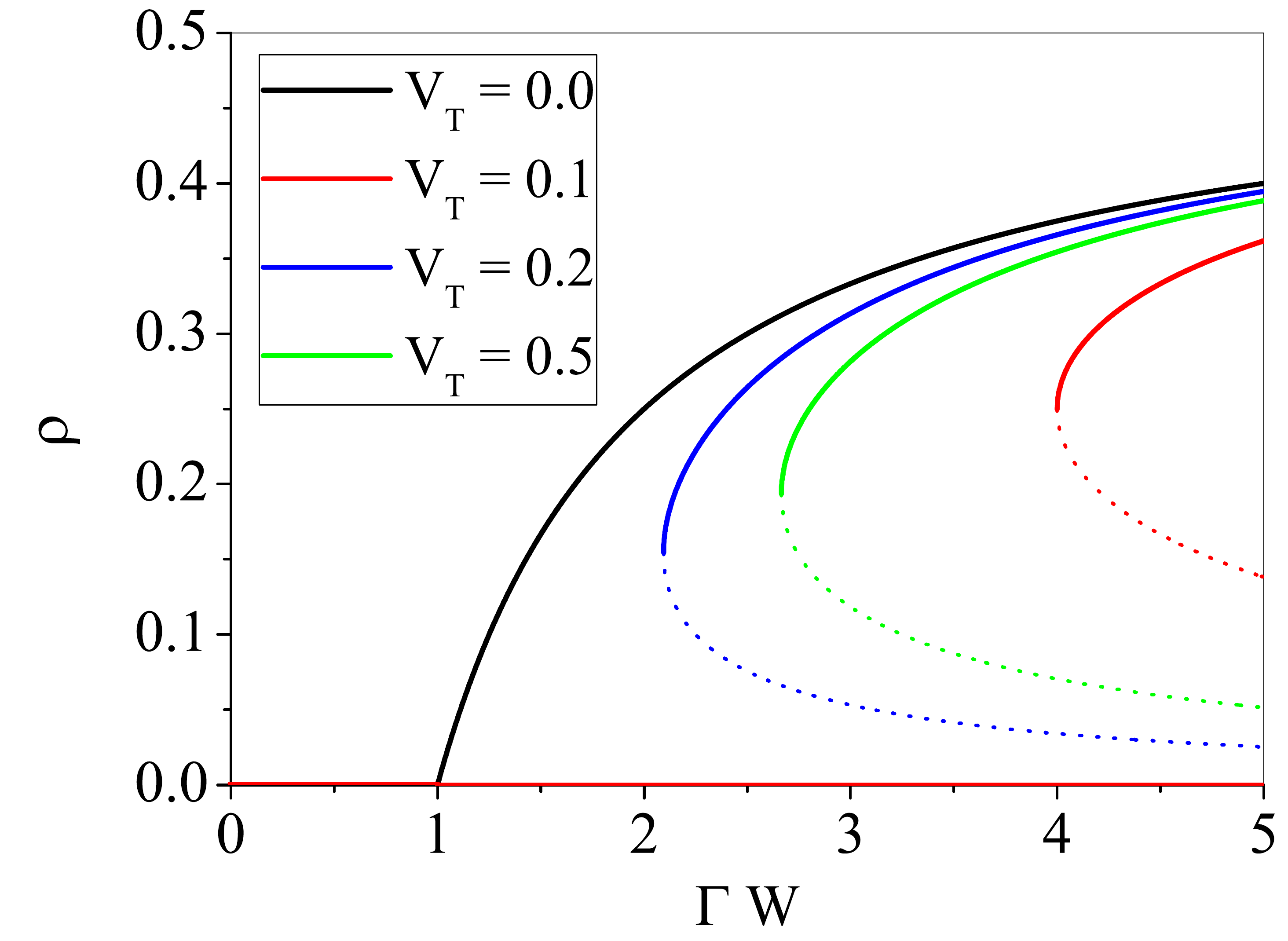}
  \caption{}
  \label{Fig2a}
\end{subfigure}
\begin{subfigure}{.45\textwidth}
  \centering
  \includegraphics[width=1.0\linewidth]{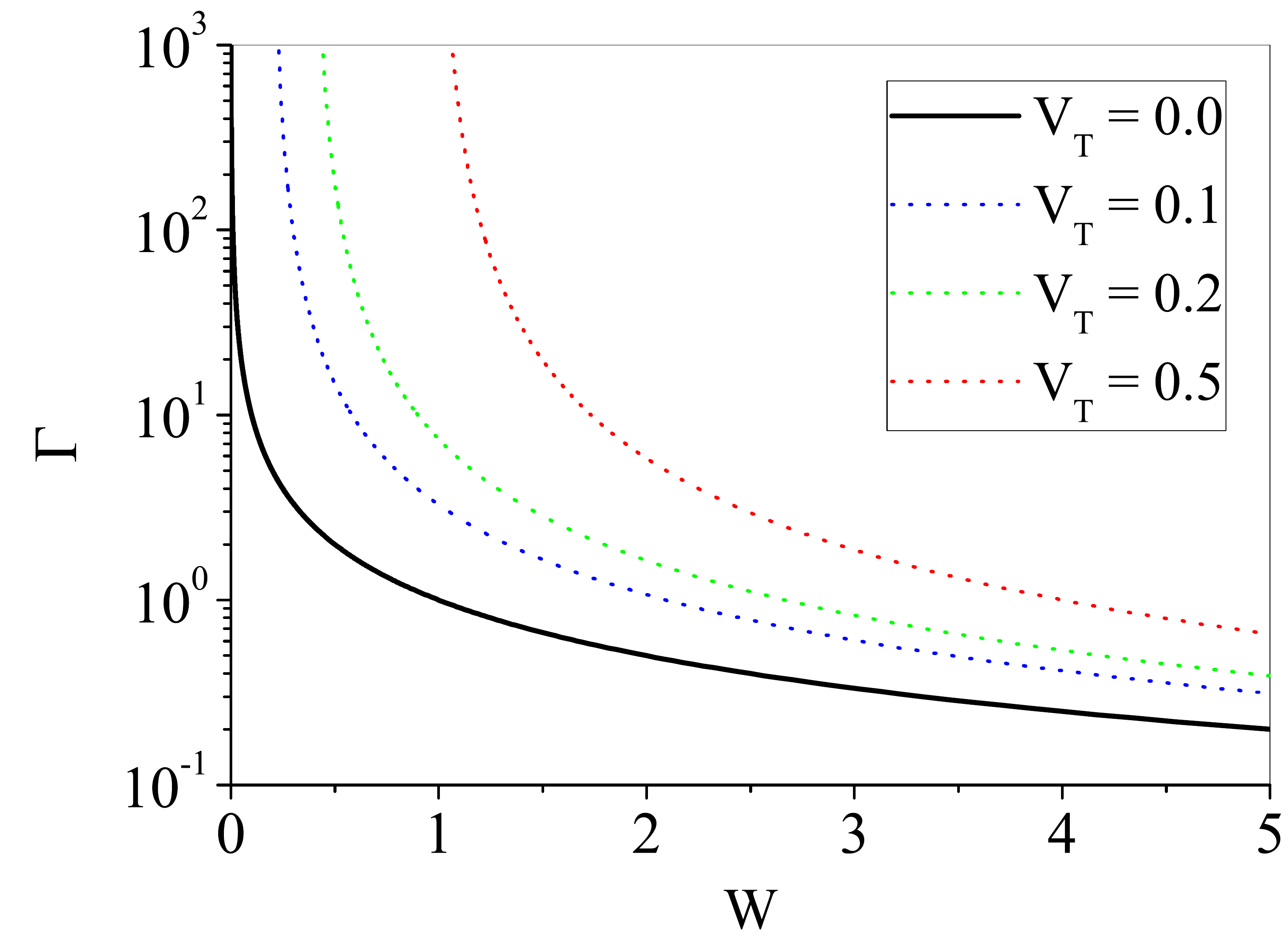}
  \caption{}
  \label{Fig2b}
\end{subfigure}
 \caption{{\bf Phase transitions for the $\mu = 0$
 case as a function of  $\Gamma, W$ and $\Vfire$ 
 with $ I=0$:} a) The solid lines represent the stable fixed 
 points $\rho^+(W)$ and dashed lines represent unstable fixed points
 $\rho^-(W)$, for thresholds $\Vfire=0.0, 0.1$, $0.2$ and $0.5$. 
 The discontinuity $\rhocrit$ given by
 Eq.~(\ref{rhocritD}) goes to zero for 
 $\Vfire \rightarrow 0$. b) Phase diagram $\Gamma \times W$
 defined by Eq.~(\ref{GW}). From top to bottom,
 $\Vfire = 0.0, 0.1$, $0.2$ and $0.5$. 
 We have $\rho^+ > 0$ above the phase transition lines. 
 For $\Vfire > 0$, all the transitions are discontinuous.
}   
\label{Vf}
\end{figure}

\begin{figure}[!ht]
  \centering  
 \makebox[\textwidth][c]{ \includegraphics[width=0.55\linewidth]
 {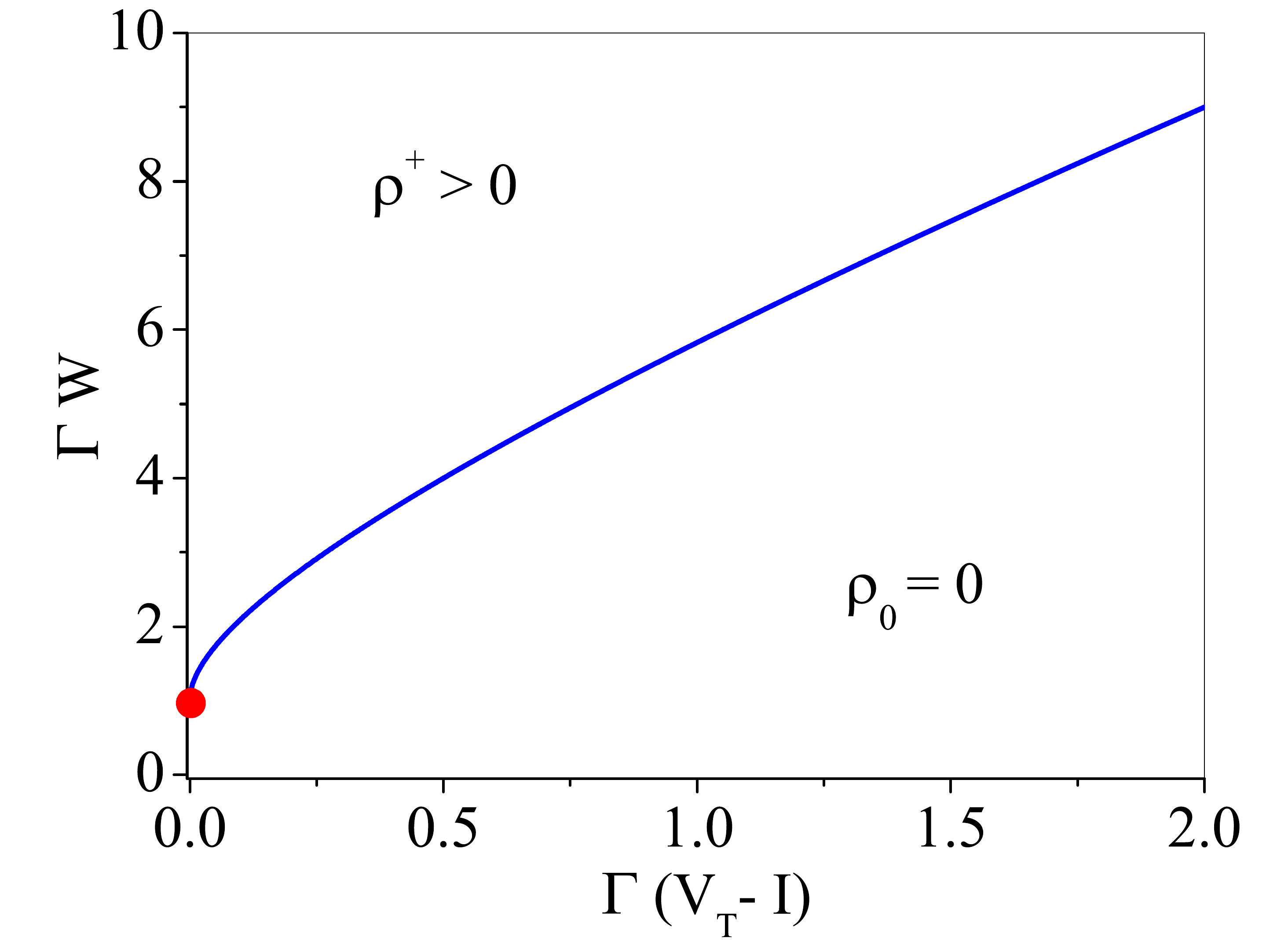}}
 \caption{{\bf Phase diagram for the $\mu = 0$
 case as a function of $\Gamma W$ and $\Gamma (\Vfire-I)$:} 
 The transition line, Eq.~(\ref{GW}), 
 is $\gcrit \Wcrit = (1+\sqrt{2\Gamma(\Vfire - I)})^2$. This line is a
 first order phase transition which terminates at the 
 second order critical point $\gcrit \Wcrit = 1$ with $\Vfire - I = 0$.}
\label{gwxvt}
\end{figure}

In Fig.~\ref{Fig2a} we show examples of the phase transitions,
which occurs when the unstable point $\rho^-$ collapses
with the stable point $\rho^+$. It is
important to notice that, for any $\Vfire >0$, 
the unstable point never touches the absorbing point $\rho_0=0$,
so the zero solution is always stable. Only for the case $\Vfire=0$ the
solution $\rho_0$ looses stability and $\rho^+$ is the unique
solution above the critical line $\gcrit = 1/ \Wcrit$.
Fig.~\ref{Fig2b} gives the phase diagram $\Gamma \times W$
for some values of $\Vfire$ with $I = 0$, see Eq.~(\ref{GW2}).
Finally, we give the phase diagram for the variables 
$ \Gamma W$ versus $\Gamma(V-I)$, see Fig.~\ref{gwxvt}
and Eq.~(\ref{GW}).

\subsection{Self-organized supercriticality (SOSC) 
through dynamic gains with  {$\mu=0$}, $I=0$, $V_T=0$}

If we fine tune the model to some point in the critical line $\gcrit = 1/ W$, we can observe perfect neuronal avalanches with size distribution $P_S(s) \propto s^{-3/2}$ and duration distribution $P_D(d) \propto d^{-2}$~\cite{Brochini2016}.
As expected, these are mean-field exponents
fully compatible with the experimental
results~\cite{Beggs2003,Hesse2015}.

This fine tuning, however, is not plausible biologically.
What we need is some homeostatic mechanism that makes
the critical region an attractor of some self-organization 
dynamics.  {In the literature, a well studied mechanism is
dynamic synapses $W_{ij}[t]$~\cite{Levina2007,Levina2009,
Bonachela2010}.}
For example, in discrete time \cite{Costa2015,Campos2017}:
\begin{equation}
W_{ij}[t+1] = W_{ij}[t] + \frac{1}{\tau} (A - W_{ij}[t] ) 
- u W_{ij}[t] X_j[t]\:,
\label{syndyn}
\end{equation}
where $\tau$ is a synaptic recovery time, $A$ is an asymptotic
value and $u \in [0,1]$ is the fraction of depletion of 
neurotransmitter vesicles when the presynaptic neuron fires.

In Brochini \emph{et al.}~\cite{Brochini2016}, we proposed a
new self-organization mechanism based in dynamic neuronal gains
$\Gamma_i[t]$ while keeping the synapses $W_{ij}$
fixed~\cite{Brochini2016}. The idea is to create a feedback loop 
based only in the local activity $X_i[t]$ of the neuron, reducing 
the gain when the neuron fires, and  recovering slowly after that.
  {The biological motivation for dynamic gains is spike frequency
adaptation, a well-known phenomenon that 
depends on the decrease (and recovery) of
sodium ion channels density at the axon initial segment 
(AIS) when the neuron fires}~\cite{Ermentrout2001,Benda2003}.

The dynamics for the neuronal gains studied in~\cite{Brochini2016} 
has a form similar to that used in 
\cite{Levina2007,Bonachela2010,Costa2015,Campos2017}
for synapses:
\begin{equation}
 \Gamma_i[t+1] = \Gamma_i[t] +\frac{1}{\tau} (A - \Gamma_i[t]) 
 - u \Gamma_i[t] X_i[t]\:.
\end{equation}

The advantage of neuronal gains is that now we have only $N$
dynamical equations (notice the term $X_i[t]$ that refers to
the activity of the postsynaptic neuron,  {not of the presynaptic
one as in Eq.~(\ref{syndyn})). For dynamic synapses, we need to simulate
$N(N-1)$ equations for the fully connected graph model, and 
$10^4 N$ for a biologically realistic network, and this is computationally very costly for large $N$.}

A problem with this dynamics, however, also present in
dynamic synapses, is that we have a three dimensional
parameter space ($\tau \in [1,\infty], A \in [1/W,\infty], u\in[0,1]$)
that must be fully explored to characterize the stationary value
$\Gamma^*(\tau,A,u,N)$.
Here we propose a new simplified dynamics with a single free
parameter, the gain recovery time $\tau$:
\begin{equation}
 \Gamma_i[t+1] = \Gamma_i[t] +\frac{1}{\tau} 
 \Gamma_i[t] - \Gamma_i[t] X_i[t] = \left( 1+ \frac{1}{\tau} - X_i[t] 
 \right) \Gamma[t]\:.
\label{gt}
\end{equation} 

\begin{figure}[!ht]
 \centering
  \begin{subfigure}{.5\textwidth}
  \centering
  \includegraphics[width=1.0\linewidth]{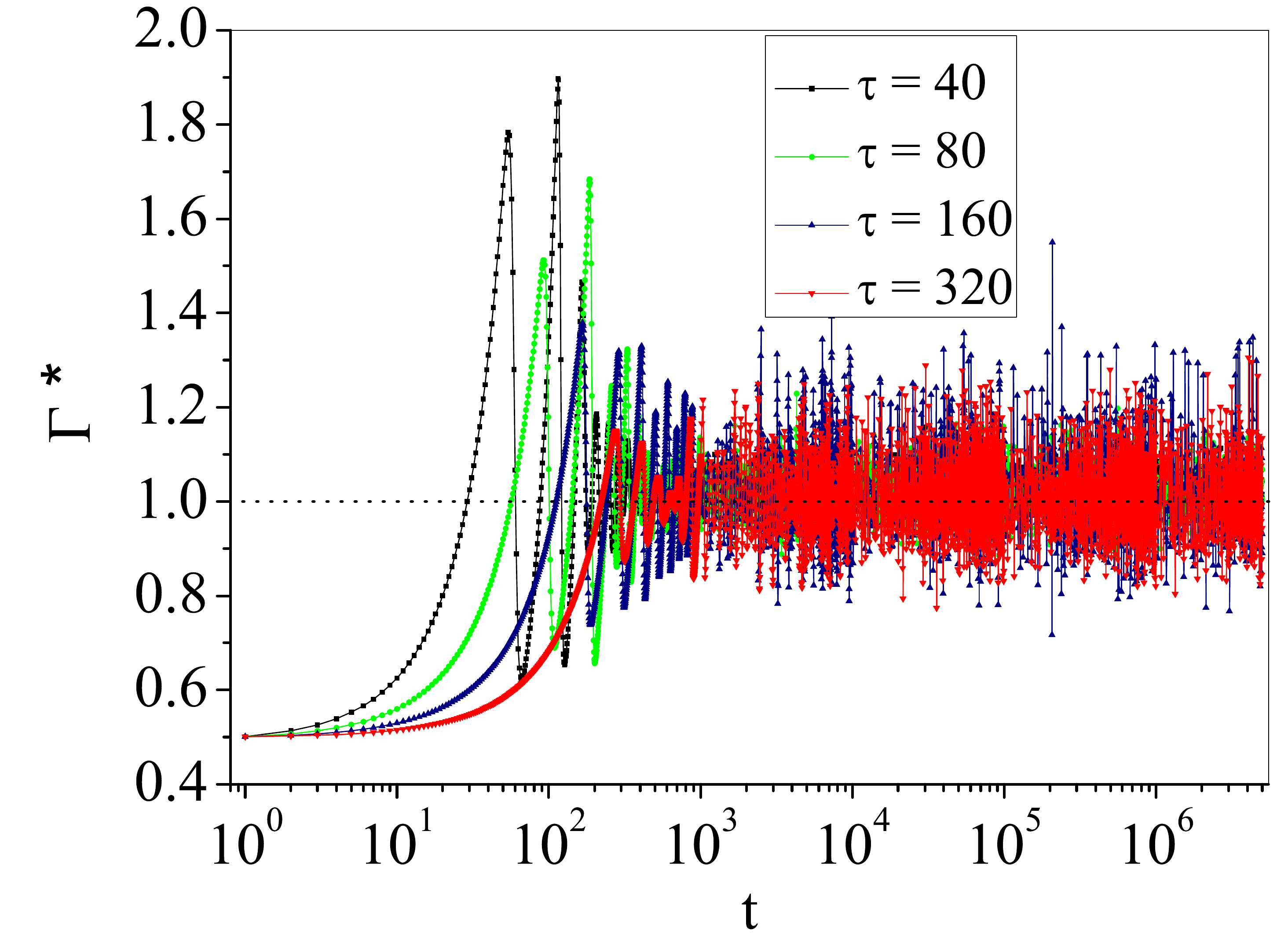}
  \caption{}
  \label{fig:sfig1}
\end{subfigure}%
\begin{subfigure}{.5\textwidth}
  \centering
  \includegraphics[width=1.0\linewidth]{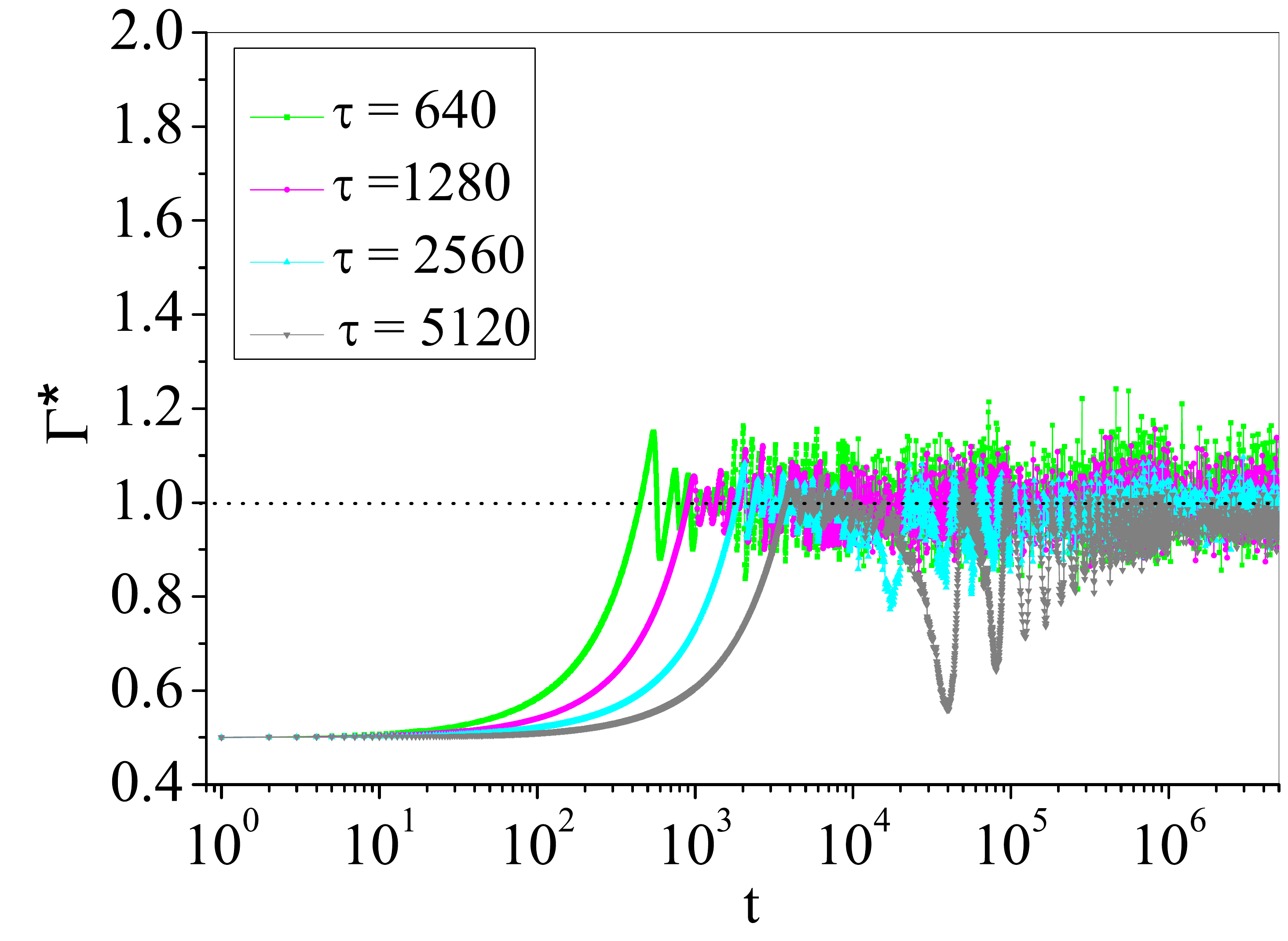}
  \caption{}
  \label{fig:sfig2}
\end{subfigure}
  
 \caption{{\bf Self-organization with dynamic neuronal gains:}
  Simulations of a network of $N=160000$ neurons with fixed 
  $W_{ij}= W = 1$ and $\Vfire = 0$. Dynamic 
  gains $\Gamma_i[t]$ starts with $\Gamma_i[0]$ uniformly 
  distributed in $[0,\Gamma_{\max}=1.0]$. This defines the
  initial condition $\Gamma[0] \equiv  \frac{1}{N} 
  \sum_i^N \Gamma_i[0] \approx \Gamma_{\max}/2 = 0.5$.
  Self-organization of the average 
  gain $\Gamma[t]$ over time, for different $\tau$. The horizontal 
  dashed line marks the value $\gcrit=1/W =1$.} 
\label{GAMMA}
\end{figure}

\begin{figure}[!ht]
 \centering
  \begin{subfigure}{.5\textwidth}
  \centering
  \includegraphics[width=1.0\linewidth]{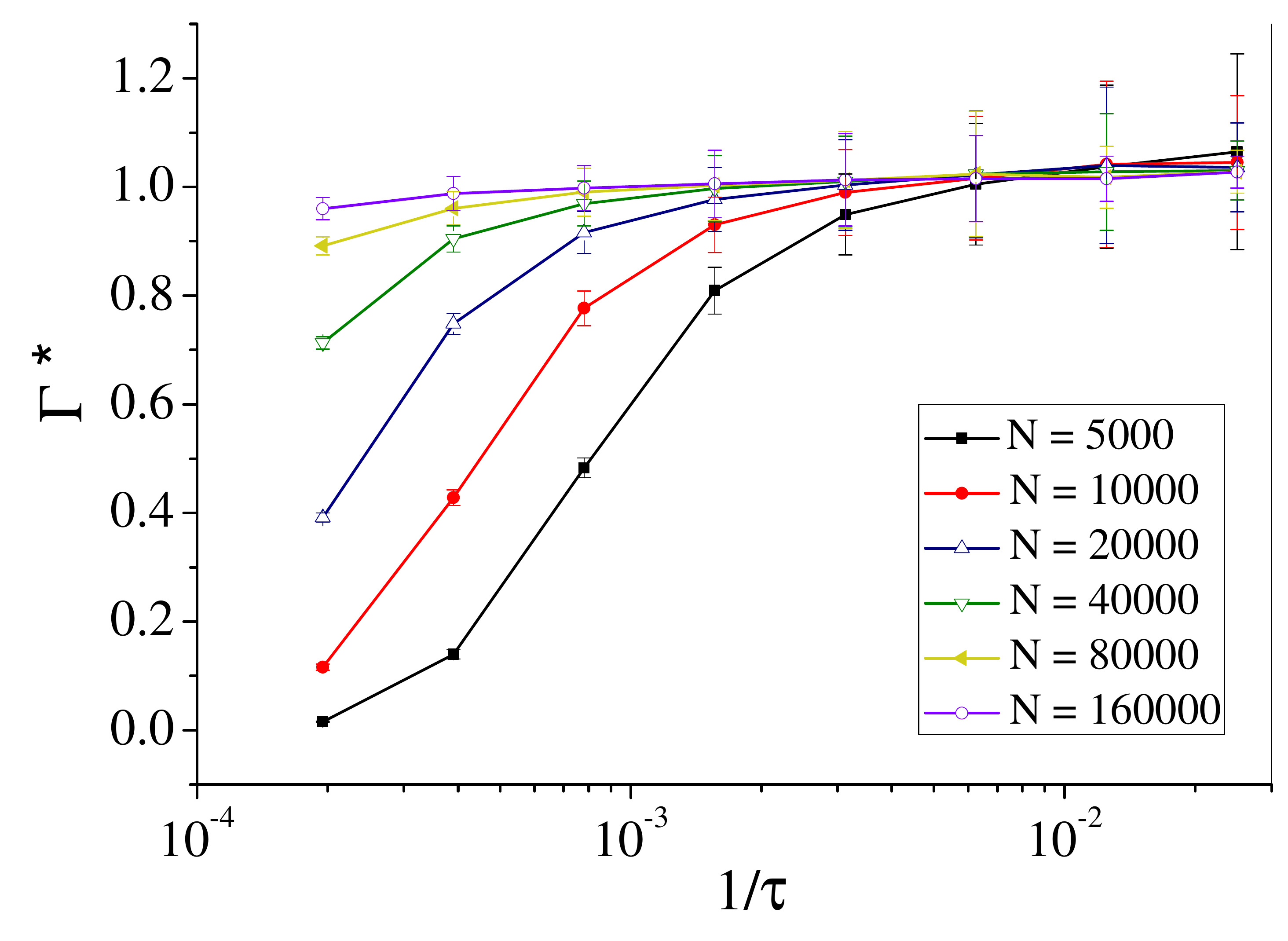}
  \caption{}
  \label{fig:gfig1}
\end{subfigure}
   \begin{subfigure}{.5\textwidth}
  \centering
  \includegraphics[width=1.0\linewidth]{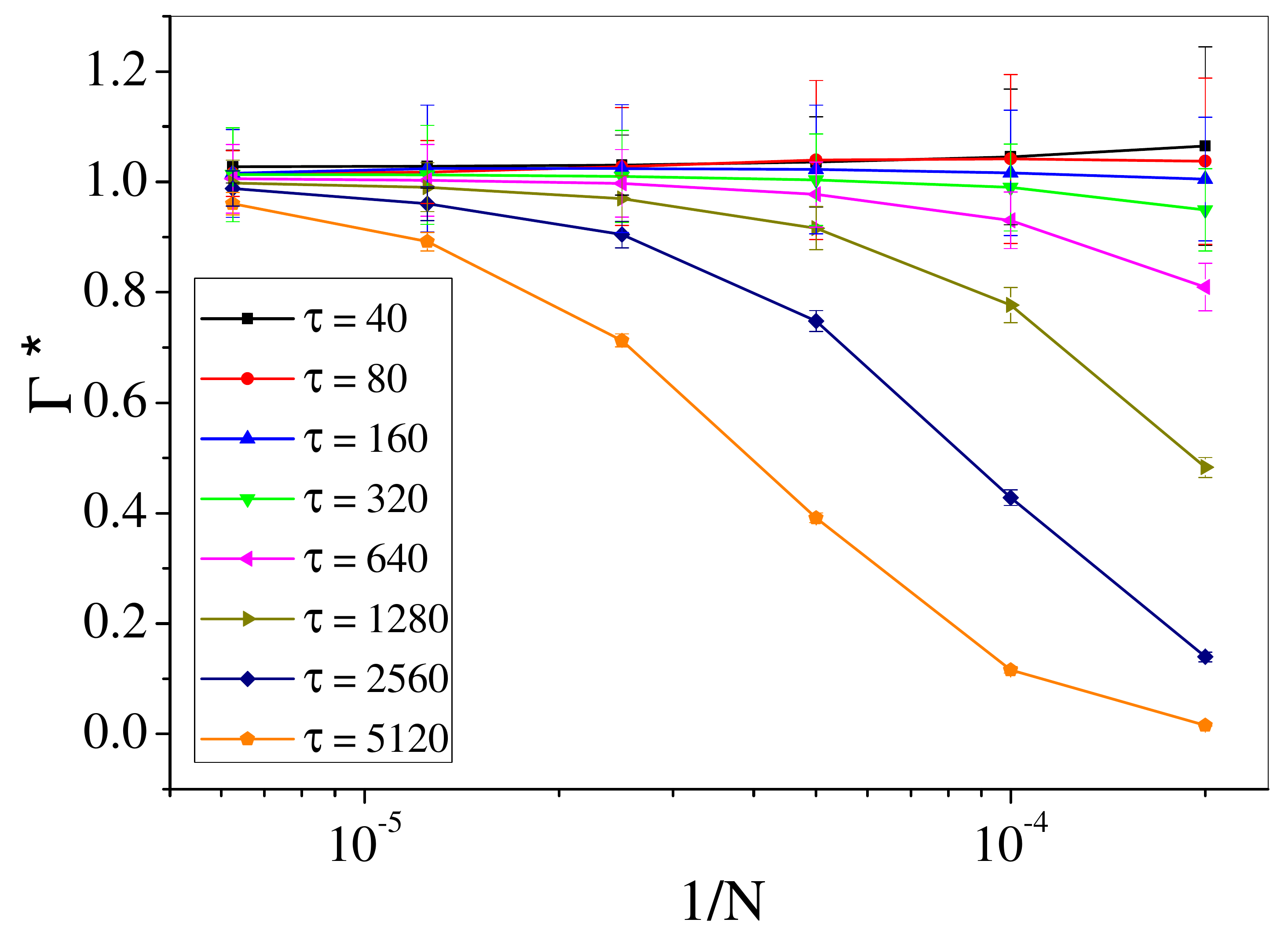}
  \caption{}
  \label{fig:gfig2}
  \end{subfigure} \\
\begin{subfigure}{.5\textwidth}
  \centering
  \includegraphics[width=1.0\linewidth]{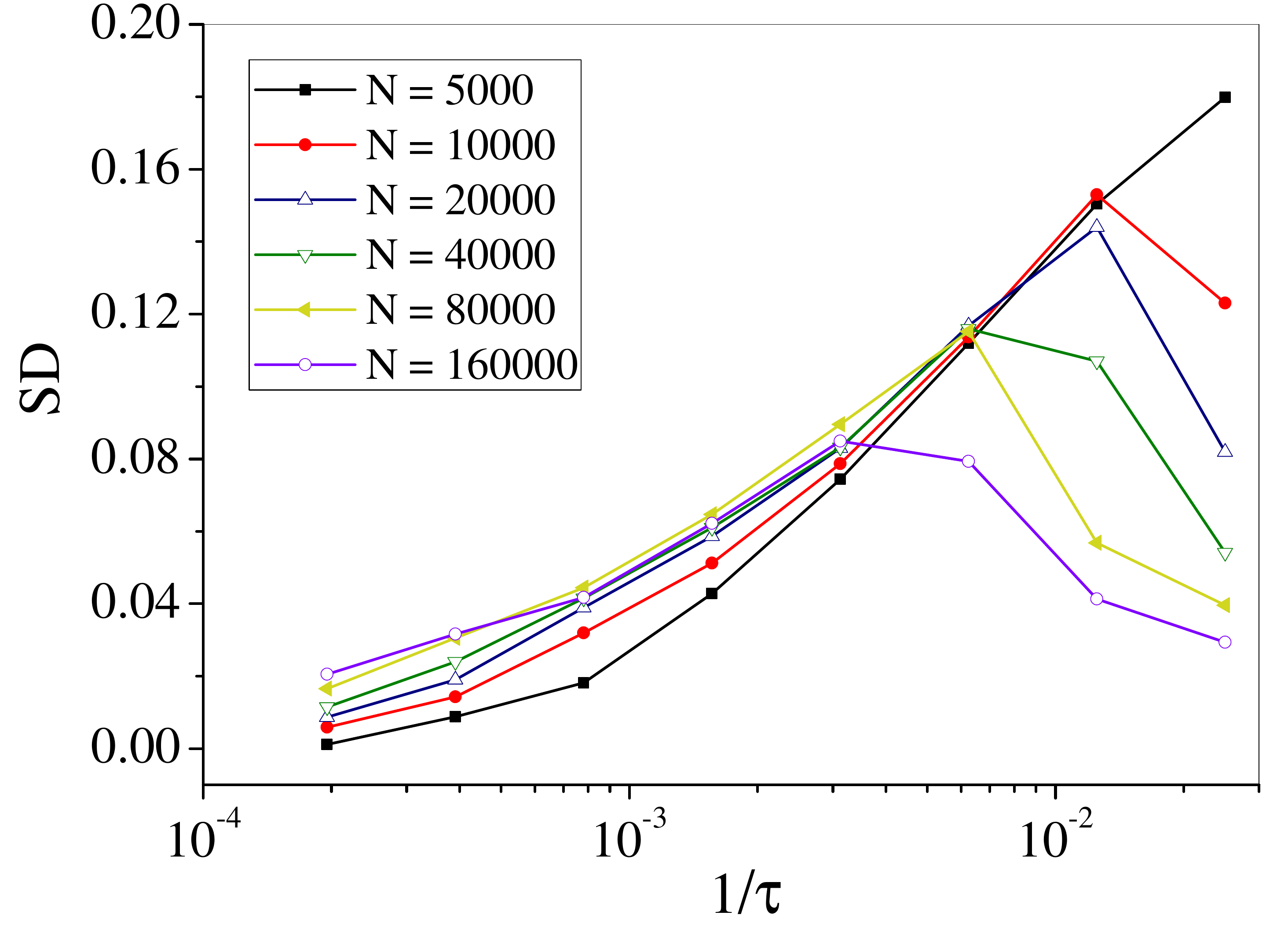}
  \caption{}
  \label{fig:gfig3}
\end{subfigure}
\begin{subfigure}{.5\textwidth}
  \centering
  \includegraphics[width=1.0\linewidth]{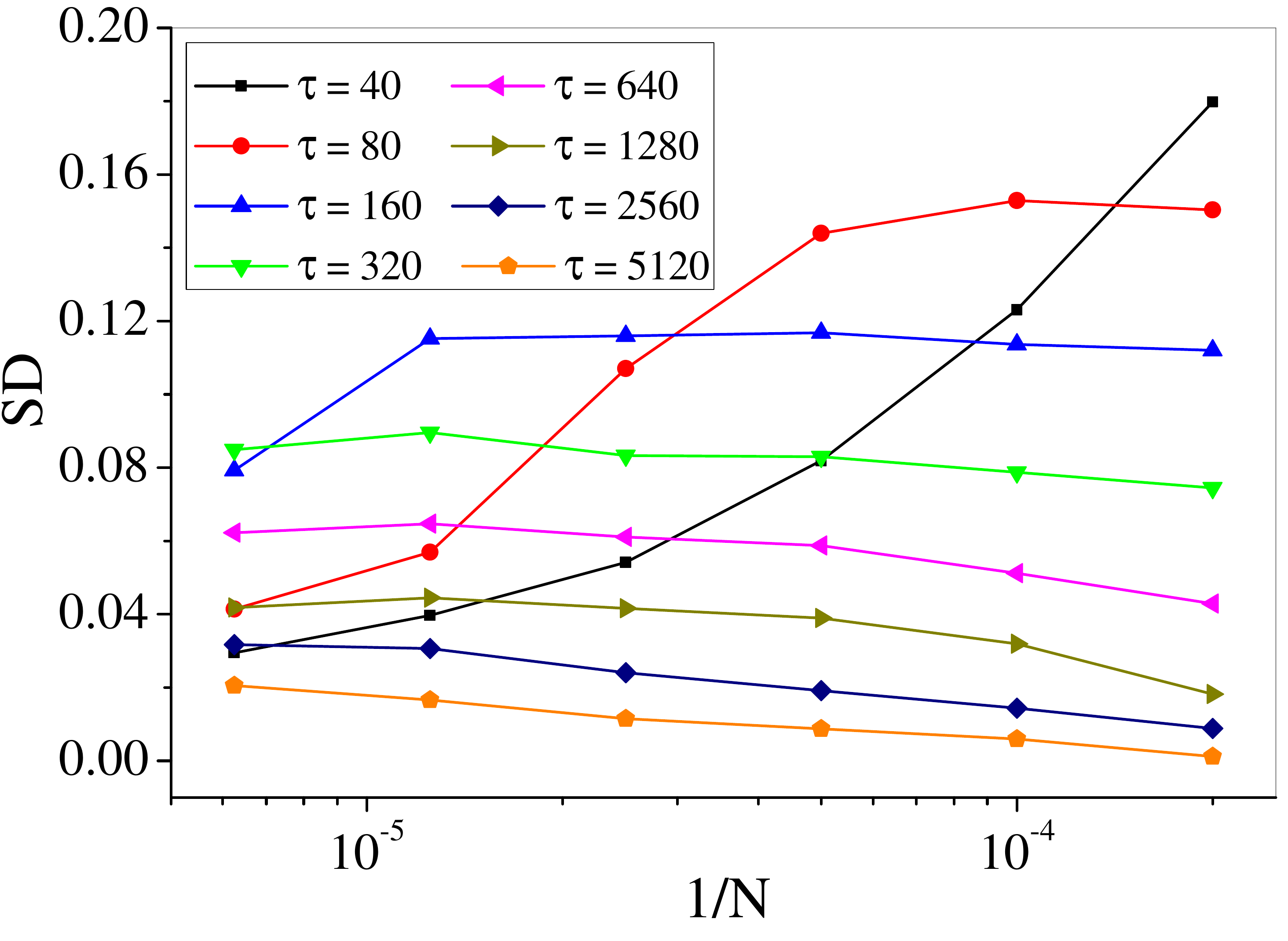}
  \caption{}
  \label{fig:gfig4}
\end{subfigure}
 
 \caption{{\bf Self-organized value $\Gamma^*(\tau,N)$ obtained with
 dynamic gains ($W_{ij} = W = 1$):} 
 a)  Curves $\Gamma(1/\tau)$ for several values of $N$.
 b) Curves $\Gamma(1/N)$ for several values of $\tau$.
 c) Standard deviation of the $\Gamma[t]$ time series after
 the transient, as a function of $1/\tau$. 
 d) Standard deviation of the $\Gamma[t]$ time series 
 after the transient, as a function of $1/N$.} 
\label{Gxtau1}
\end{figure}

The self-organization mechanism can be viewed in Fig.~\ref{GAMMA}.
So, we reduce our parametric study to determine the curves
$\Gamma^*(1/\tau,1/N)$, see Figs.~\ref{fig:gfig1} and~\ref{fig:gfig2}. 
The fluctuations measured by the standard deviation $SD$
of the $\Gamma[t]$ time series, after the transient, 
diminishes for increasing $\tau$~(Fig.~\ref{fig:gfig3}) and 
probably goes to zero for $\tau \rightarrow \infty$, in accord to
Campos \emph{et al.}~\cite{Campos2017}. However, 
in contrast to this idealized $\tau \rightarrow \infty$ limit, 
as discussed in the references~\cite{Bonachela2010,Campos2017},
the fluctuations do not converge to zero for finite $\tau$ in the
thermodynamic limit $N \rightarrow \infty$~(see Fig.~\ref{fig:gfig4}).
 {This occurs because, for low $\tau$, the adaptation mechanism
produces oscillations of $\Gamma[t]$ around the value $\Gamma^*(\tau)$}

We can do a mean-field analysis of Eq.~(\ref{gt})  {to find the
value $\Gamma^*(\tau)$.} 
Denote the average gain as $\Gamma[t] = \avg{\Gamma_i[t]}$. 
Averaging over the sites, we have : 
\begin{equation}
\Gamma[t+1] = \Gamma[t] +\frac{1}{\tau}\Gamma[t] 
- \rho[t] \:\Gamma[t] \:,
\end{equation}
since $\rho[t] = \avg{X_i[t]}$.
In the stationary state, we have $\Gamma[t+1] = 
\Gamma[t] = \Gamma^*,\: \rho[t] = \rho^*$, so:
\begin{equation}
\frac{1}{\tau}\:\Gamma^* = \rho^* \:\Gamma^*\:. \label{gamma*}
\end{equation}

A solution is $\Gamma^* = 0$, but this is unstable, see Eq.~(\ref{gt}).
Another solution is obtained  by inserting Eq.~(\ref{rhocrit}),
$\rho^* =(\Gamma^* - \gcrit)/(2 \Gamma^*)$, in Eq.~(\ref{gamma*}):
\begin{equation}
\Gamma^* =  \frac{\gcrit}{1 - 2/\tau}  \:. \label{gamma**}
\end{equation}
Notice that this is valid only when 
Eq.~(\ref{rhocrit}) is valid, that is, for $\Gamma^* \geq 1/W$. 
 {Also, Eq.~(\ref{gamma**}) presumes that
$\rho^* $ is a stable fixed point, which can
not be true for some interval of values of
$\tau$, see below.}

A first order approximation leads to:
\begin{equation}
\Gamma^* = \gcrit \left(1 + \frac{2}{\tau} \right) \:.
\label{gamma2}
\end{equation}
This mean-field calculation shows that, if $\tau \rightarrow 
\infty$, we obtain an exact SOC state $\Gamma^* \rightarrow \gcrit$.
Or, for finite networks, it would be required a scaling 
$\tau = O(N^a)$ with an exponent $a>0$, as done previously for dynamic 
synapses~\cite{Levina2007,Bonachela2010,Costa2015,Campos2017}.
However, this scaling for $\tau$ cannot be justified biologically.

\begin{figure}[!ht]
  \centering
  \makebox[\textwidth][c]{\includegraphics[width=\linewidth]{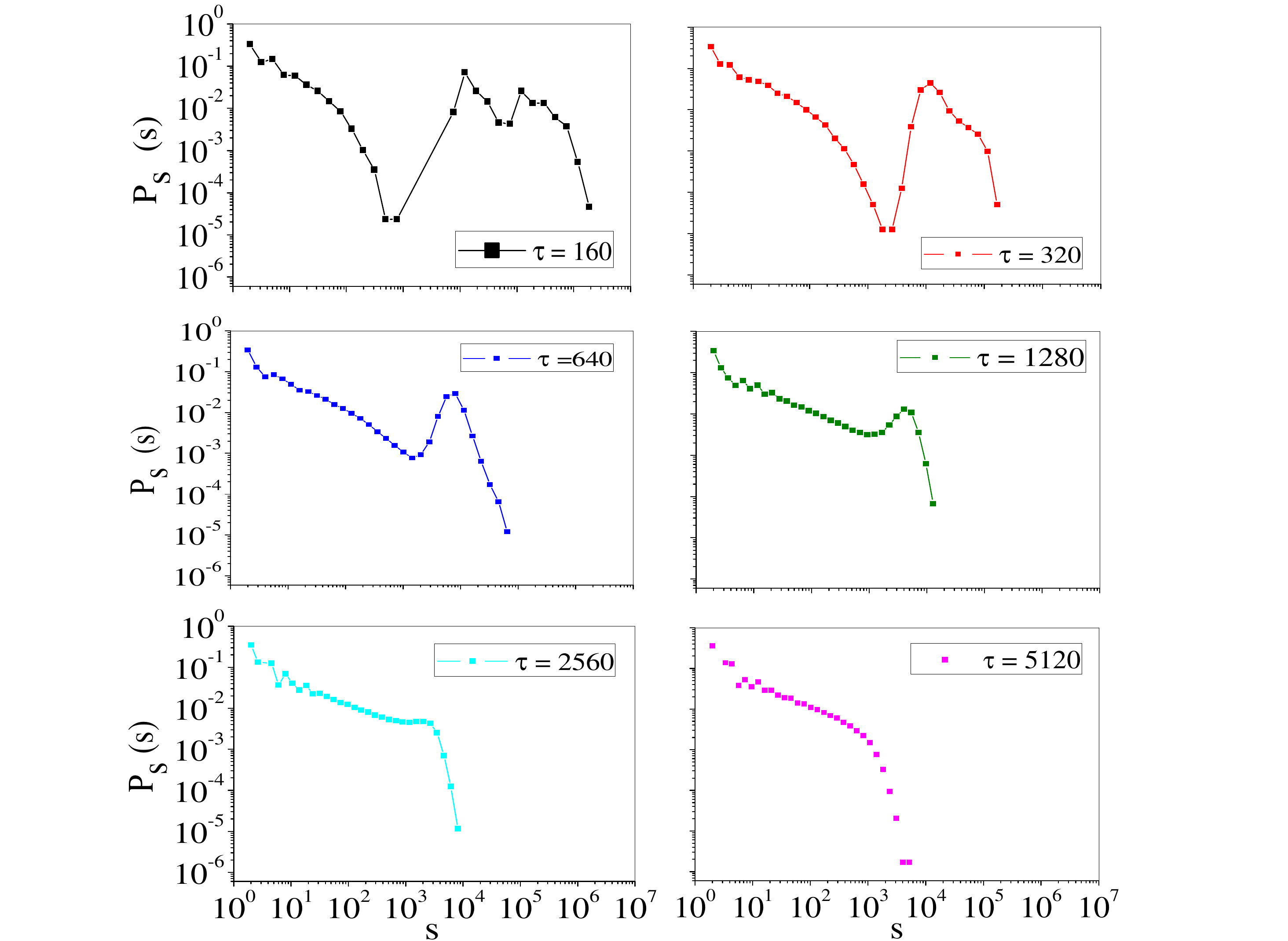}}
  \caption{{\bf Avalanche statistics for the model
  with dynamic neuronal gains:} Probability histogram for
  avalanche sizes ($P_S(s))$  {with logarithmic bins} for several $\tau$ with $N = 160000$.
  Notice the SOSC phenomenon and Dragon-king avalanches for
 small $\tau$.} 
\label{Avalanches}
\end{figure}

So, biology requires a finite recovery time $\tau$ which
always leads to supercriticality, see Eq.~(\ref{gamma**}) or (\ref{gamma2}).
This supercriticality is self-organized in the sense that it is 
achieved and maintained  by the gain dynamics Eq.~(\ref{gt}).
We call this phenomena self-organized supercriticality (SOSC).

The deviation from criticality can be small.
For example, if $\tau = 1000$ ms (assuming 1 time step 
equals 1 ms in the model):
\begin{equation}
\Gamma^* \approx 1.002 \:\gcrit \:. 
\end{equation}
Even a more conservative value $\tau = 100$ ms gives 
$\Gamma^* \approx 1.02 \: \gcrit$.
Although not perfect SOC~\cite{Markovic2014}, 
this result is  sufficient to explain 
a power law with exponent $3/2$ for small ($s < 1000$) neuronal 
avalanches plus a supercritical bump (Fig.~\ref{Avalanches}).

By using Eq.~(\ref{gamma**}) in Eq.~(\ref{rhocrit}) we also obtain:
\begin{equation}
\rho^* = \frac{1}{\tau}  \:
\end{equation}
 { {showing that the network presents supercritical activity for any finite $\tau$.}}
This result, however, is valid only in the infinite size limit.
For finite size networks, fluctuations interrupt this constant
$\rho^* = 1/\tau$ activity, leading the system to the absorbing 
state and defining the end of the avalanches.

\begin{figure}[!ht]
 \centering
   \begin{subfigure}{.48\textwidth}
  \includegraphics[width=1.0\linewidth]{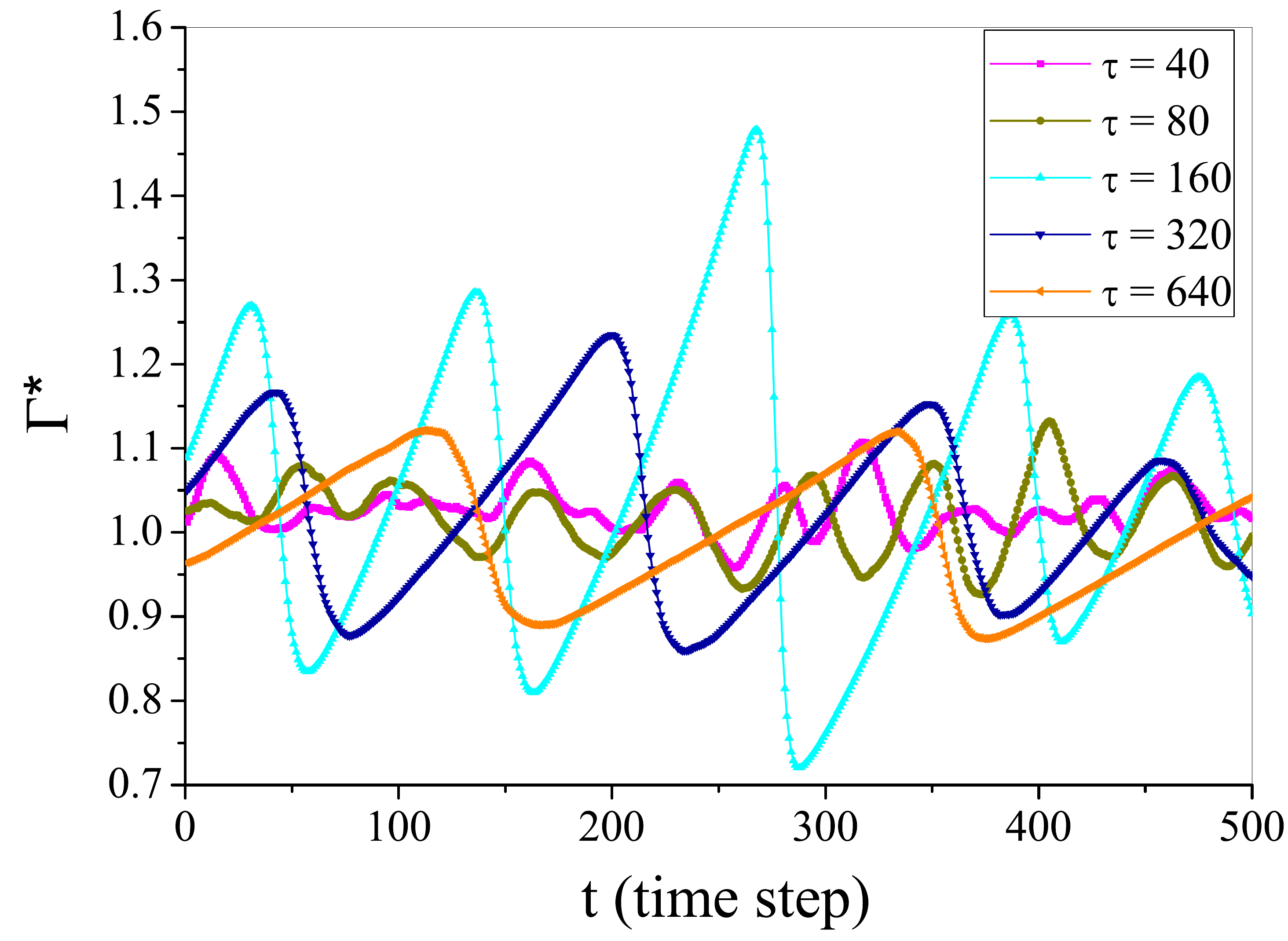}
  \caption{}
  \label{fig7a}
\end{subfigure}
\begin{subfigure}{.48\textwidth}
  \includegraphics[width=1.0\linewidth]{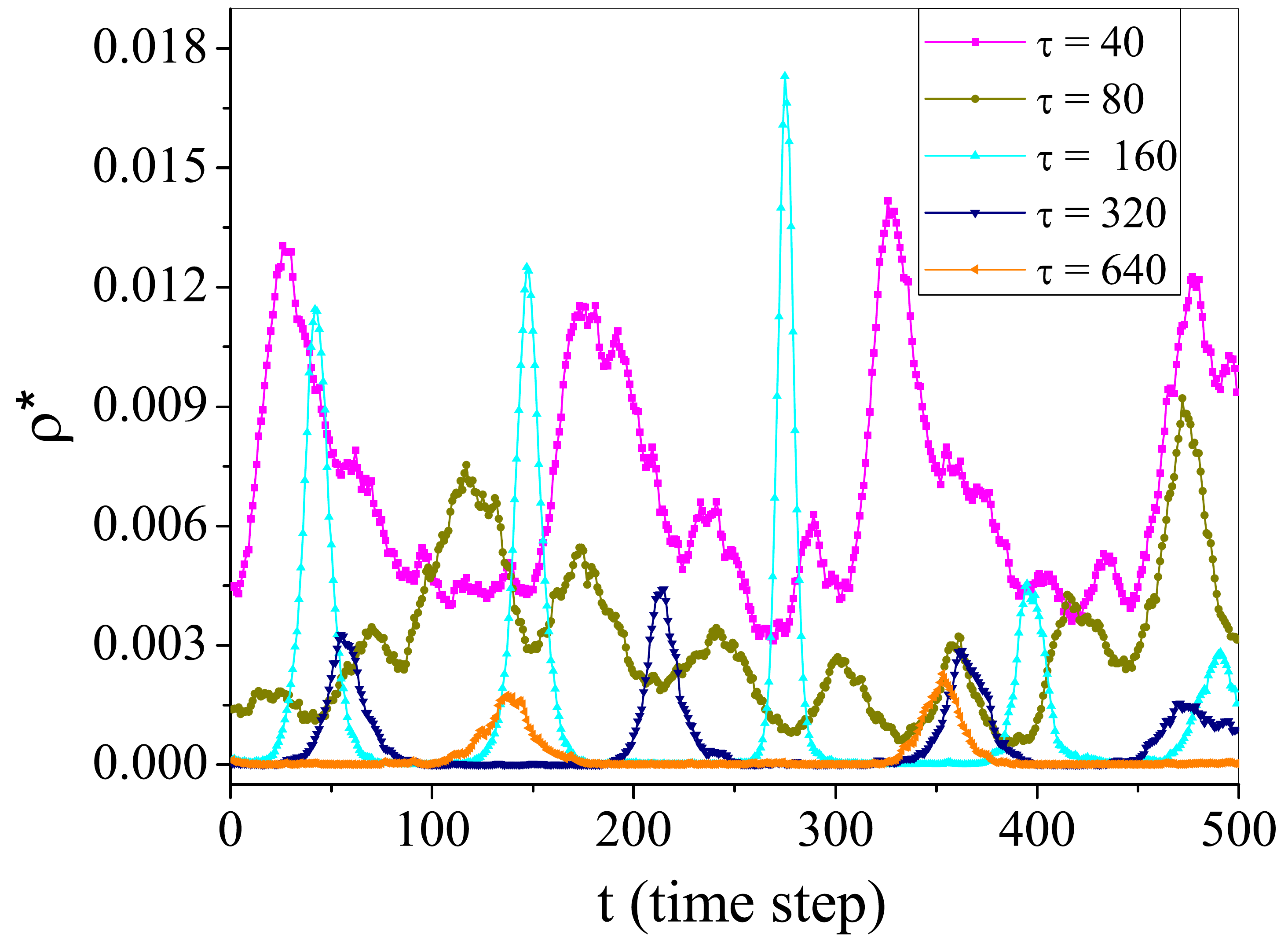}
  \caption{}
  \label{fig7b}
\end{subfigure}   \\
    \begin{subfigure}{.48\textwidth}
  \centering
  \includegraphics[width=1.0\linewidth]{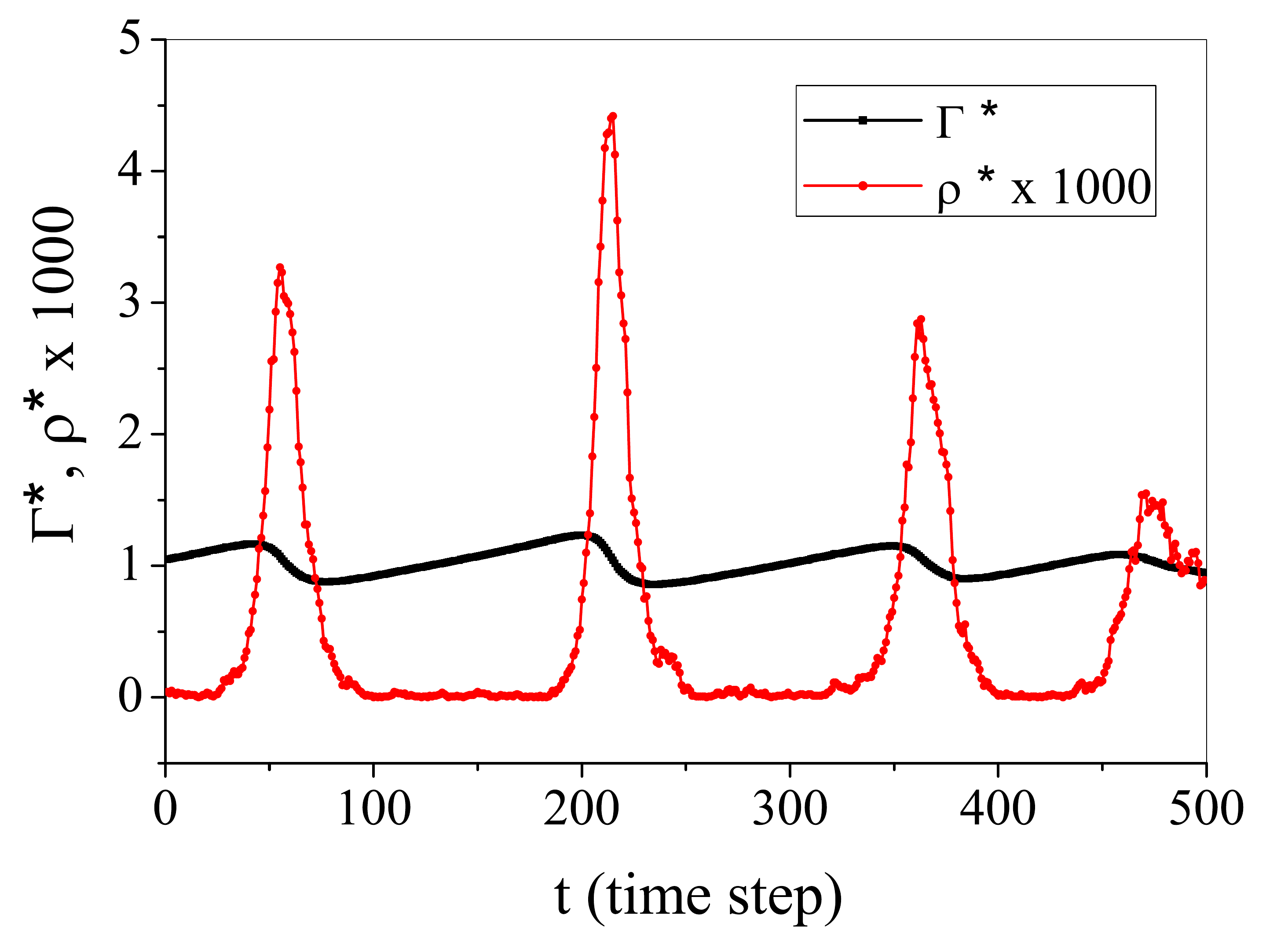}
  \caption{}
  \label{fig7c}
\end{subfigure}
    \caption{{
    \bf Dynamic gain $\Gamma[t]$ and activity $\rho[t]$.} 
      (a) and (b)$\Gamma[t]$ for several values of $\tau$; 
      (c) $\Gamma[t]$ and $\rho[t]$ for $\tau = 320$. 
      In the figures we consider only the last 500 time steps of simulation (from a time series of 5 million time steps)  in a system with $N=160,000$.  {The large events (oscillations) correspond to Dragon-king avalanches.}
  }
\label{figgammat}
\end{figure}

 {Simulations reveal that these fixed points  {$(\Gamma^*, \:\rho^*)$} correspond only to mean values around which both 
$\Gamma[t]$ and $\rho[t]$ oscillate}, see Figs.~\ref{figgammat}a,
\ref{figgammat}b and \ref{figgammat}c. 
These global oscillations are unexpected since the model has been
devised to produce avalanches, not oscillations. 
The finite size fluctuations and oscillations drive the
network to the absorbing zero state, generating the avalanches. 
What we see in the histogram of Fig.~\ref{Avalanches} is a 
combination of power law avalanches in some range plus
large events  {(superavalanches or Dragon-kings due to the $\Gamma[t]$ oscillations).}

\section{Discussion}

We examined a network of stochastic spiking neurons with a rational
firing function $\Phi$ that has not been studied previously. 
We obtained numeric and analytic results that show the presence of continuous and discontinuous
absorbing phase transitions. Classic SOC is possible only at the
continuous transition, which means that we need to use zero
firing thresholds ($\Vfire =0$). In some sense, this is a 
kind of fine tuning of the $\Phi$ function,
but not the usual one where the synaptic strength $W$ and 
the neuronal gain $\Gamma$ are the main control
parameters.

The presence of a well behaved absorbing phase 
transition in the directed percolation class enables the use of
an homeostatic mechanism for the neuronal gains that tunes
the network to the critical region. 
 {The dynamics on the gains is biologically plausible, and can be 
related to a decrease and recovery, due to the neuron activity,
of the firing probability at the Axon Initial Segment
(AIS)~\cite{Kole2012}. Our dynamic $\Gamma_i[t]$ mimics the 
well known phenomenon of \emph{spike frequency
adaptation}~\cite{Ermentrout2001,Benda2003} and
is a one-parameter simplification of the three-parameter dynamics studied by us in~\cite{Brochini2016}.}

 {We observe that this gain dynamics is equivalent to approach the
critical line with fixed $W$ and variable $\Gamma[t]$, that is,
performing vertical movements in Fig.~\ref{Fig2b}. Previous
literature approaches the critical point $\Wcrit$
by dynamic~\cite{Levina2007,Levina2009,
Bonachela2010,Costa2015,Campos2017} or Hebbian~\cite{Arcangelis2006,Pellegrini2007,Arcangelis2010,Arcangelis2012,
Arcangelis2012b,Kessenich2016} synapses. This corresponds
to fix $\Gamma$ and allow changes in $W_{ij}[t]$
along the horizontal axis, see Fig.~\ref{Fig2b}.} 

 {The two homeostatic strategies are similar, but we stress
that we have only $N$ equations for the gains $\Gamma_i[t]$ instead
of $N(N-1)$ equations for the synapses $W_{ij}[t]$, so that our
approach imply a huge computational advantage. Indeed, previous literature
as~\cite{Levina2007,Arcangelis2012} report system sizes on the
range of $N = 1,000 - 4,000$, to be compared to our maximal size of
$N = 160,000$.} 

We found that  the fixed point $\Gamma^*$ predicted 
by a mean-field calculation is not exactly critical, 
but instead supercritical, and that the distance
from criticality depends on the gain recovery time $\tau$. 
Previous claims about achieving exact SOC by using dynamic synapses
are based on the erroneous assumption that we can use
a synaptic recovery time 
$\tau \propto N^a \rightarrow \infty$~\cite{Levina2007,Bonachela2010,
Costa2015,Campos2017}. But if we use a finite $\tau$, which is not only  plausible but biologically necessary,
we obtain SOSC, not SOC~\cite{Campos2017,Brochini2016}. 
 {However, we found that for large but
plausible values of $\tau$, the system
is only slightly supercritical and presents power law 
avalanches (plus small supercritical bumps) 
compatible with the biological data.}

SOSC enables us to explore supercritical
networks that are robust, that is, the stationary state,
with or without oscillations, is achieved from any initial
condition and recovers from perturbations. So, the question now is:
are there  self-organized supercritical (SOSC) oscillating
neuronal networks in the brain?

A first evidence would be a supercritical bump in the 
$P(s)$ distributions. Indeed, we found several papers where
such bumps seem to be present, see for example the first plot in
Fig.~2 of Friedman \emph{et al.}~\cite{Friedman2012} and Fig.~4 of  
Scott \textit{et al.}~\cite{Scott2014}.
It seems to us that, since the main paradigm for neuronal avalanches
is exact SOC, with pure power laws, 
 { {it is possible that}} researchers report what is expected
and do not comment or emphasize small supercritical bumps, even if they  are present in their published data.  Therefore, we suggest that experimental researchers reevaluate
their data in search for small supercritical bumps. The presence of supercritical bumps can also be masked by the phenomenon of subsampling~\cite{Priesemann2009,Girardi2013,
Levina2017}, so the analysis must be done with some care. 

Supercriticality, in the form of the so called Dragon-king
avalanches~\cite{Sornette2012,Arcangelis2012,Lin2017}, has been conjectured to 
be at the basis of hyperexcitability in
epilepsy~\cite{Hesse2015,Hobbs2010,Meisel2012}.
Also, networks can be put artificially in the hyperexcitable state
and show bimodal distributions $P_S(s)$ with large 
supercritical bumps~\cite{Haldeman2005}.
The SOSC phenomenon seems to be a natural explanation for
such hyperexcitability. In~\cite{Haldeman2005}, the supercritical
bumps are fitted by a supercritical branching process but are
not explained in mechanistic terms as, in our case, due to
different values of the biophysical $\tau$ recovery time. 

The unexpected oscillations 
in $\Gamma[t]$ around $\Gamma^*$ have 
amplitudes that depend on $\tau$ and vanish
for large $\tau$ (Fig.~\ref{fig:gfig4} and Fig.~\ref{fig7a}).
These oscillations in $\Gamma[t]$ induce oscillations 
in the activity $\rho[t]$ (Fig.~\ref{fig7b} and \ref{fig7c}). 
In our model, the discrete time
interval $\Delta t$ is postulated as describing 
the width of a spike, that is, $\Delta t \approx 1-2$ ms. 
From our simulation data, with these values
for $\Delta t$, we obtain frequencies 
$f \approx 0.5 - 16$ Hz, depending
on the $\tau$ value. 

Interestingly, this frequency range includes
Delta, Theta and Alpha rhythms. Coexistence  of Theta
waves  and neuronal avalanches has been 
observed experimentally~\cite{Gireesh2008}. 
Also, some theoretical  work
recently  discussed  coexistence of
oscillations and avalanches~\cite{Poil2012}.

The presence of oscillations
can mean that the fixed point $(\Gamma^*, \rho^*)$ is unstable
below some biffurcation point $\tau_b$
(even for $N\rightarrow \infty$) or that it is stable but 
has a very small negative Lyapunov exponent, such that finite-size
fluctuations drive $\Gamma[t], \rho[t]$ away from equilibrium, producing
excursions (oscillations) in the $(\Gamma, \rho)$ plane.
At this point, without further study,
we can not decide what is the correct scenario.  Notice that
similar oscillations for $W[t]$ were also observed  for 
dynamic synapses~\cite{Levina2007,Bonachela2010,Costa2015},
although these authors have not studied in detail such phenomenon.

Finally, from a conceptual point of view, the
observed subcriticality in some of our simulations (see 
Fig.~\ref{fig:gfig2} and \ref{fig:gfig4})  is less 
important than supercriticality (SOSC), because it is 
a finite-size effect for small
$N$. Our largest networks have $N = 160,000$, which is small
compared to real biological networks that have at least one or
two orders of magnitude more neurons. 

Nevertheless, there is in the literature claims 
that subcritical states are present in certain experimental 
conditions~\cite{Bedard2006,Tetzlaff2010,Priesemann2014}. How
we can conciliate these findings? Here we offer an answer based on the findings of Priesemann \emph{et al.}~\cite{Priesemann2014}.
These authors found that, in order of explain  \emph{in vivo} experiments with awake animals, they need three ingredients: subsampling~\cite{Priesemann2009}, increased input (violating the standard separation of scales of SOC  models) and small subcriticality of the networks. If we increase the inputs in our network, by a Poisson process on the variable $I[t]$ for example, the overall result is that the homeostatic mechanism turns out our network subcritical. This occurs because increased forced firing imply an overall depression of the gains $\Gamma_i[t]$ in Eq.~(\ref{gt}), 
so that a new equilibrium is achieved with 
$ \Gamma^* <  \Gamma_c$. 

Then, under external input like in
awake animals, our adaptive networks turns out subcritical and
returns to criticality or supercriticality for spontaneous
activity without external input.
We obtained preliminary simulation results confirming this scenario and a comprehensive
study of the effect of external input shall be done in a next paper.

\section{Materials and Methods}


All numerical calculations were done by using MATLAB.
Simulation codes were made in Fortran90. 

In the study of neuronal avalanches, we simulate the evolution of 
finite networks with $N$ neurons, uniform synaptic strengths 
$W_{ij} = W$ ($W_{ii} = 0$) and $\Phi(V)$ rational with
$\Vfire=0$.
The avalanche statistics were obtained after the transient
of the neuronal gains
self-organization. A silent instant when
$X_i[t]=0$ for all $i$ defines the end of an avalanche. 
We start a new avalanche by forcing the firing of 
a single random neuron $i$, setting $V_i[t+1]$ to a value high enough for the neuron spikes. 

\section{Conclusion}

We have shown in this paper that dynamic neuronal gains
lead naturally to self-organized supercriticality (SOSC) and not
SOC. The same occurs with dynamic synapses~\cite{Campos2017}.
So, we propose that neuronal avalanches are related to SOSC instead
of exact SOC. This opens an opportunity for reevaluation of the
accumulated  experimental data. 

SOSC suggests that neuronal tissues
could be more prone to Dragon-king avalanches~\cite{Sornette2012} and
hyperexcitability than one would
expect from simple power laws. This prediction of larger and 
increased instability due to supercriticality 
may be important for studies in epilepsy~\cite{Arcangelis2012}.

Finally, the  emergence of oscillations coexisting with 
neuronal avalanches seems to unify in a single formalism two theoretical approaches and two different research communities: those that emphasize critical behavior and avalanches, and those that emphasize oscillations and synchronized activity.

In a future work, we intend to study with more care
the mechanism that generates these oscillations and how to relate them to EEG data. In order to simulate more biological networks, we also intend to study the cases $V_T > 0, I>0$ and $\mu>0$.

\vspace{6pt} 


\acknowledgments{
This article was produced as part of the activities of FAPESP  Research, Innovation and Dissemination Center for Neuromathematics (grant \#2013/07699-0, S.Paulo Research Foundation). The RIDC for Neuromathematics covered publication costs. AAC also thanks grants $\#$2016/00430-3 and $\#$2016/20945-8 São Paulo Research Foundation (FAPESP).
LB also acknowledges grant $\#$2016/24676-1 São Paulo Research Foundation (FAPESP). OK also received support from N\'ucleo de Apoio \`a Pesquisa CNAIPS-USP.
}

\authorcontributions{AAC performed the network simulations, and prepared all the figures. 
OK and LB performed analytic and numerical calculations. All authors analyzed the
results and wrote the manuscript.}

\section*{Conflicts of interest:} The authors declare no conflict of interest.



\appendix

\section{Phase transition for $\mu>0, \Vfire = 0$}

We want to derive the critical point for the leakage case $\mu > 0$ and
also to obtain approximate curves for the activity $\rho$ near the
critical region. We start from the exact formulas 
(supposing $\Phi(0) = 0$):
\begin{eqnarray}
\rho &=& \sum_{k=1}^\infty \eta_k \Phi(U_k) \:, \label{A1}\\
\eta_k &=& \eta_{k-1}\left(1-\Phi(U_{k-1}) \right) \:, \label{A2}\\
U_0 &=& 0 \:,   \label{A3}\\
U_k &=& \mu U_{k-1} + W \rho \:,\label{A4} \\
&=& W \rho \sum_{j=0}^{k-1} \mu^j = W\rho \frac{1-\mu^k}{1-\mu} \:,\label{A5} \\
\end{eqnarray}
then, we use the recurrence relations Eq.~(\ref{A2}) and (\ref{A4}) into Eq.~(\ref{A1}):
\begin{eqnarray}
\rho &=& \sum_{k=1}^\infty \eta_{k-1}\left(1-\Phi(U_{k-1})\right)
\Phi(\mu U_{k-1} + W\rho)\:. \label{A6}
\end{eqnarray}

We notice that, due to Eq.~(\ref{A4}), all terms $U_k$ are small
in the critical region where $\rho \rightarrow 0$.
So, we approximate the rational $\Phi(U)$ function for small $U$:
\begin{equation}
\Phi(U_k) = \frac{\Gamma U_k}{1+\Gamma U_k} 
\approx \Gamma U_k  - \Gamma^2 U_k^2 \:. \label{proxphi}
\end{equation}

Inserting in Eq.~(\ref{A6}), and using Eq.~(\ref{A4}) we get:
\begin{eqnarray}
\rho &=&\sum_{k=1}^\infty \eta_{k-1} (1- \Gamma U_k  + \Gamma^2 U_k^2) (\Gamma \mu U_{k-1} + \Gamma W \rho - \Gamma ^2(\mu U_{k-1} + W \rho)^2)\:.
\end{eqnarray}

Since each $U_k$ is proportional to $\rho$, 
from now we conserve only terms
proportional to $\rho$ and $\rho^2$. After recombining the terms up to order $V_{k-1}^2$ according to Eq. (\ref{proxphi}), we obtain:

\begin{eqnarray*}
\rho & \approx & \Gamma W \rho (1 - \Gamma W \rho) \sum_{k=1}^\infty \eta_{k-1} 
+ (\mu -2\Gamma \mu W \rho -\Gamma W \rho + \Gamma^2 W^2 \rho^2) \sum_{k=1}^\infty \eta_{k-1} \phi(U_{k-1}) \\
&-& \mu^2\Gamma \sum_{k=1}^\infty \eta_{k-1} \phi(U_{k-1}) U_{k-1}\:.
\end{eqnarray*}

Notice that $\sum_{k=1}^\infty \eta_{k-1} =
\sum_{k=0}^\infty \eta_{k} =1$ by normalization. Using this fact and also Eq. (\ref{A1}) (using $\Phi(U_0) = 0$), after some rearrangement we obtain:
\begin{eqnarray}
\rho\approx \rho(\Gamma W + \mu)+\rho^2(-\Gamma^2 W^2-2\Gamma W \mu - \Gamma W ) -  \mu^2 \Gamma \sum_{k=1}^\infty \eta_{k-1} \phi(U_{k-1}) U_{k-1}\:. \label{Arho1}
\end{eqnarray}

With respect to the last term, we use Eq. (\ref{A5}) and  (\ref{A1}) to obtain
\begin{eqnarray*}
 \mu^2 \Gamma \sum_{k=1}^\infty \eta_{k-1} \phi(U_{k-1}) U_{k-1}
&=&\mu^2 \Gamma \sum_{k=1}^\infty \eta_{k-1} \phi(U_{k-1}) (W\rho\frac{1-\mu^k}{1-\mu})\\
&=&\frac{\mu^2 \Gamma W \rho}{1-\mu} \left( \rho -\sum_{k=1}^\infty \eta_{k-1} \phi(U_{k-1}) \mu^k \right)\\
&\approx& \frac{\mu^2 \Gamma W \rho^2}{1-\mu}\:,
\end{eqnarray*}
which is valid for $\mu<1$. Here, the sum $\rho \sum_{k=1}^\infty \eta_{k-1} \phi(U_{k-1}) \mu^k $ is composed of terms in $\rho^3$  than can be dismissed. Using this approximation in Eq. \ref{Arho1} we obtain two solutions. One is the absorbing state $\rho=0$. The other solution is
\begin{eqnarray}
\rho \approx \frac{1}{1+\Gamma W+ 2\mu + \mu^2/(1-\mu) } \: \frac{\Gamma -\Gamma_c}{\Gamma}\:, 
\end{eqnarray}
where we considered $\Gamma_c=(1-\mu)/W$. Moreover, in the critical region we can approximate $\Gamma W \approx 1-\mu$, leading to:
\begin{eqnarray}
\rho\approx  \frac{1}{2+\mu+ \mu^2/(1-\mu) } \: \frac{\Gamma -\Gamma_c}{\Gamma}\:.
\end{eqnarray}

We compare this analytical approximation with numerical solutions for 
$\rho(\Gamma W,\mu)$ near the critical point, see Fig.~\ref{Fig1c}.

A similar calculation for the monomial function
$\Phi(V) =\Gamma V \Theta(V) \Theta(\Gamma V-1) +\Theta(1-\Gamma V)$
gives:
\begin{equation}
\rho \approx (1-\mu) \: \frac{\Gamma-\Gamma_c}{\Gamma} \:.
\end{equation}
with the same critical line $\gcrit = (1- \mu)/W$. The
monomial function with $\mu > 0$ was studied numerically in Brochini 
\emph{et al.}~\cite{Brochini2016} but this analytic proof
for $\gcrit$ is new.

\bibliographystyle{mdpi}

\renewcommand\bibname{References}
\bibliography{bibsuper}

\end{document}